\documentclass{aa}  
\usepackage{graphicx}
\usepackage{natbib}
\usepackage{hyperref}
\usepackage[varg]{txfonts}
\usepackage{xcolor}

\begin{document} 
  \title{Properties of the Hyades, the eclipsing binary HD\,27130, and the oscillating red giant $\epsilon$\,Tau}
  
  \author{K. Brogaard
          \inst{1, 2}
          \and
          E. Pak{\v s}tien{\. e}
          \inst{2}
          \and
          F. Grundahl
          \inst{1, 2}
                    \and
          \v{S}. Mikolaitis
          \inst{2}
                    \and
          G. Tautvai{\v s}ien{\. e}
                    \inst{2}
                              \and
          D. Slumstrup
                    \inst{3, 1}
                              \and\\
          G. J. J. Talens
                    \inst{4}
                          \and
          D. A. VandenBerg
                    \inst{5}
                              \and
        A. Miglio \inst{6}
                              \and
          T. Arentoft
                    \inst{1}
                              \and
          H. Kjeldsen
                    \inst{1, 2}
                    \and
          R. Janulis
                    \inst{2}
                              \and
          A. Drazdauskas
                    \inst{2}
                              \and\\
          A. Marchini   
                    \inst{7}
                              \and
          R. Minkevi{\v c}i{\= u}t{\. e}
                    \inst{2}
                              \and
           E. Stonkut{\. e}
                    \inst{2}
                              \and     
          V. Bagdonas
                    \inst{2}
  \and     
          M. Fredslund Andersen
                    \inst{1}
                      \and     
          J. Jessen-Hansen
                    \inst{1}
                      \and     
          P. L.  Pallé
                    \inst{8,9}
                    \and                
          P. Dorval
                    \inst{10,11}
                    \and
          I. A. G. Snellen
                    \inst{10}
                    \and
          G. P. P. L. Otten
                    \inst{12}
                    \and
          T. R. White
                    \inst{13,1}
         }

  \institute{Stellar Astrophysics Centre, Department of Physics and Astronomy, Aarhus University, Ny Munkegade 120, DK-8000 Aarhus C, Denmark
   \and
  Astronomical Observatory, Institute of Theoretical Physics and Astronomy, Vilnius University, Saul\.{e}tekio av. 3, 10257 Vilnius, Lithuania
  \and
  European Southern Observatory, Alonso de Córdova 3107, Vitacura, Santiago, Chile
   \and
  Institut de Recherche sur les Exoplan\`{e}tes, D\'{e}partement de Physique, Universit\'{e} de Montr\'{e}al, Montr\'{e}al, QC H3C 3J7, Canada
  \and
  Department of Physics and Astronomy, University of Victoria, Victoria, BC, V8W 2Y2, Canada
  \and
  School of Physics and Astronomy, University of Birmingham, Birmingham B15 2TT, United Kingdom
  \and
  Astronomical Observatory, Department of Physical Sciences, Earth and Environment - University of Siena, Via Roma 56, 53100 Siena, Italy
  \and
  Instituto de Astrofísica de Canarias, E-38200 La Laguna, Tenerife, Spain
  \and
  Universidad de La Laguna (ULL), Departamento de Astrofísica, E-38206 La Laguna, Tenerife, Spain
  \and
  Leiden Observatory, Leiden University, Postbus 9513, 2300 RA Leiden, The Netherlands
  \and
  NOVA Optical IR Instrumentation Group at ASTRON, P.O. Box 2, 7990 AA Dwingeloo, The Netherlands
  \and
  Aix Marseille Univ, CNRS, CNES, LAM, Marseille, France
  \and
  Sydney Institute for Astronomy (SIfA), School of Physics, University of Sydney, NSW, 2006, Australia
  }

  \date{Received XXX / Accepted XXX}

  \abstract{
    Eclipsing binary stars allow derivation of accurate and precise masses and radii, from which insights
       into their evolution may be obtained.  When they reside in star
       clusters, properties of even higher precision, along with
       additional information, can be extracted. Asteroseismology of solar-like oscillations offers similar possibilities for single stars.
}{
    We wish to improve the previously established properties of the Hyades eclipsing binary HD\,27130 and re-assess the asteroseismic properties of the giant star $\epsilon$ Tau. The physical properties of these members of the Hyades are then used to constrain the helium content and age of the cluster.
}{
    New multi-colour light curves were combined with multi-epoch radial velocities to yield masses and radii of HD\,27130. $T_{\rm eff}$ was derived from spectroscopy and photometry, and verified using the Gaia parallax.
    We estimate the cluster age from re-evaluated asteroseismic properties of $\epsilon$ Tau while using HD\,27130 to constrain the helium content.
}{
    The masses and radii, and $T_{\rm eff}$ of HD\,27130 were found to be 
    $M=1.0245\pm0.0024\,M_{\odot}$, $R=0.9226\pm0.015\,R_{\odot}$, $T_{\rm eff}=5650\pm50$\,K for the primary, and $M=0.7426\pm0.0016\,M_{\odot}$, $R=0.7388\pm0.026\,R_{\odot}$, $T_{\rm eff}=4300\pm100$\,K for the secondary component. 
    Our re-evaluation of $\epsilon$ Tau suggests that the previous literature estimates are trustworthy, and that the Hipparcos parallax is more reliable than the Gaia DR2 parallax. 
}{
    The helium content of HD\,27130 and thus of the Hyades is found to be $Y=0.27$ but with significant model dependence. Correlations with the adopted metallicity results in a robust helium enrichment law with $\frac{\Delta Y}{\Delta Z}$ close to 1.2 
    
    We estimate the age of the Hyades to be 0.9 $\pm$ 0.1 (stat) $\pm$ 0.1 (sys) Gyr in slight tension with recent age estimates based on the cluster white dwarfs. The age precision can be much improved by asteroseismology of the other Hyades giants, and by future improvements to the Gaia parallax of bright stars.}

\keywords{binaries: eclipsing -- open clusters and associations: individual: Hyades
Stars: oscillations -- stars: abundances -- stars: individual (HD\,27130, $\epsilon$\,Tau)} 

\maketitle

\section{Introduction}

Open star clusters offer an opportunity to investigate stellar evolution in detail due to the shared properties of their member stars. Analysis of eclipsing binary stars and asteroseismology are strong tools to improve such studies further \citep[e.g.][]{Brogaard2011,Brogaard2012,Brogaard2017, Brogaard2018,Brogaard2018A,Miglio2012,Miglio2016,Handberg2017,Arentoft2019,Sandquist2020}. Here, we investigate the eclipsing binary HD\,27130 and the oscillating giant $\epsilon$ Tau to constrain the properties of the Hyades open cluster.

The Hyades is a  relatively young open cluster. According to the literature it is $625-790$\,Myr \citep[e.g.][]{Perryman98, Brandt15, Gossage18, Martin18,Gaia2018}, has super-solar metallicity ($\text{[Fe/H]}$ determinations range from $+0.1$\,dex to $+0.2$\,dex, \citealt{Takeda2020} and references therein), and is located at a distance from the Sun of $46.75\pm{0.46}$\,pc \citep{Gaia17}.

\object{HD\,27130} (BD\,+16 577, vB\,22, V818\,Tau, HIP 20019) is a double-lined eclipsing binary system \citep{Schiller87} in the Hyades open cluster \citep[see its membership information in][]{Griffin88, Schwan91, Perryman98, Douglas14}. According to previous studies, the components constituting the HD\,27130 binary system are G8\,V (hereafter primary) and K3-5\,V (hereafter secondary) stars \citep{Schiller87, Svechnikov04}, with a period of about 5.6 days \citep{McClure82, Svechnikov04, Watson06, Peterson87, Peterson88}. 

\citet{Schiller87} carried out the $UBVRI$ observations of HD\,27130 and used the light curves to obtain the effective temperatures of $5470$\,K and $3977$\,K for the primary and secondary components, respectively. Slightly higher $T_\text{eff}$ values equal to $5530\pm{100}$\,K and $4220\pm{150}$\,K were determined by \citet{Torres02}. They also derived masses of $1.08\pm{0.017}$\,$M_\odot$ and $0.771\pm{0.011}$\,$M_\odot$ as well as radii of $0.905\pm{0.019}$\,$R_\odot$ and $0.773\pm{0.010}$\,$R_\odot$ for the primary and secondary members of the system, respectively. Similar mass values of $1.064\pm{0.011}$\,$M_\odot$, $1.072\pm{0.010}$\,$M_\odot$, $1.0591\pm{0.0062}$\,$M_\odot$ for the primary and $0.763\pm{0.005}$\,$M_\odot$, $0.769\pm{0.005}$\,$M_\odot$, $0.7605\pm{0.0062}$\,$M_\odot$ for the secondary components were attributed by \cite{Peterson87, Peterson88, Torres02}, respectively. 

The observational spectroscopic data, and orbital elements of the star system were reviewed by \citet{Griffin12}, including the detection of a faint third component. The impact of this on the physical parameters of HD\,27130 seems to have gone unnoticed. Recently, \citet{Torres2019A,Torres2019B} re-examined the mass-$M_V$ relation for the Hyades based on new and updated parameters for some of the astrometric spectroscopic binaries in the cluster. Along with their new measurements, they included literature measurements for other spectroscopic binaries, including HD\,27130, for which they adopt the measurements of \citet{Torres02}. As stressed by \citet{Torres2019B}, significant improvements are to be expected from new measurements of more of these systems. Specifically, some of the systems have properties that are based on rather old measurements. \citet{Torres2019A} mention in their introduction that HD\,27130 has its masses measured to better than 1\%, presumably based on published numbers, including those of \citet{Torres02} that they adopt for their mass-luminosity relation of the Hyades. However, given the triple nature of the system, and the updated spectroscopic parameters given by \citet{Griffin12}, the previously published masses \citep{Peterson87, Peterson88, Torres02} are likely not accurate at the precision level of 1\%. 

Here, we obtain new photometric and spectroscopic measurements of HD\,27130, and measure new precise and accurate properties of the components. This is used to constrain the helium content of the Hyades open cluster.

The giant star $\epsilon$ Tau is also a member of the Hyades. Given its late evolutionary stage, it is much better suited for an age estimate than the main-sequence components of HD\,27130. \citet{Arentoft2019} showed that $\epsilon$ Tau displays solar-like oscillations and measured its asteroseismic parameters $\Delta\nu$ and $\nu_{\rm max}$. We adopt the asteroseismic measurements of \citet{Arentoft2019} and re-asses the physical properties of $\epsilon$ Tau to estimate the age of the Hyades cluster.

The paper outline is as follows. First, we present our observations of HD\,27130 in Sect.~\ref{sec:observations}. Sect.~\ref{data reduction} contains our data reduction and basic analysis, while Sect.~\ref{sec:binary} describes our eclipsing binary modelling. $\epsilon$ Tau is presented and re-assessed in Sect.~\ref{sec:epstau}. 
The properties of HD\,27130 and $\epsilon$ Tau are used in Sect.~\ref{sec:results} to establish the helium content and the age of the Hyades. Summary, conclusions, and outlook are given in Sect.~\ref{conclusions}. 

\section{Observations}
\label{sec:observations}
HD\,27130 was observed both photometrically and spectroscopically.  
Time series of photometric and spectral observations were taken at the Mol\.{e}tai Astronomical Observatory of the Vilnius University (MAO, Lithuania). Additional photometric observations of the secondary eclipse were performed at the Astronomical Observatory of the Siena University (AO SU) in Italy. We also include a light curve from the MASCARA telescope. All light curves are available as electronic tables at CDS.

Spectral observations were also obtained from the Stellar Observations Network Group (SONG) Hertzsprung telescope located at the Teide Observatory of the Astrophysics Institute of the Canary Islands, Spain.   

\subsection{Photometric observations at the Mol{\. e}tai Astronomical Observatory }

At MAO we used a 51\,cm Maksutov-type telescope with the 35\,cm working diameter of the primary mirror and the Apogee Alta U47 CCD camera. The observations were carried out in a semi-robotic mode. During the period between 16 September, 2018, and 22 October, 2019, we observed 17 runs and took in total 11906 images in the Johnson-Cousins photometric system $B$, $V$, and $I$ filters. Exposure times varied between 2.5\,s for the $I$ filter and 5\,s for the $B$ filter. In Table~\ref{table:MAO-phot}, we present information about the observing runs at MAO. 


\begin{table}
\caption{Dates of runs and numbers of images in different filters observed for HD~27130 at MAO and AO SU.}  
\label{table:MAO-phot}      
\centering                          
\begin{tabular}{c c c c }        
\hline\hline                 
Dates of runs  & $B$ images &   $V$ images   &  $I$ images	\\
(JD-2458000) & & & \\
\hline 
378.46496-378.56400	&122      & 136    	 & 134    \\
383.43266-383.56133	&120      & 119    	 & 110    	\\
393.43604-393.59516	&46       & 23     	 & 5      	\\
401.41744-401.46681	&55       & 51     	 & 54     	\\
402.41152-402.64392	&218      & 236    	 & 232    	\\
403.41291-403.64824	&362      & 368    	 & 370    	\\
404.41396-404.63789	&340      & 354    	 & 338    	\\
405.42203-405.62959	&338      & 323    	 & 286    	\\
406.42025-406.62940	&395      & 400    	 & 397    	\\
407.42510-407.62244	&372      & 366    	 & 344    	\\
409.40506-409.64799	&463      & 300    	 & 314    	\\
451.42251-451.59986	&286      & 270    	 & 259    	\\
452.42069-452.59040	&222      & 220    	 & 224    	\\
493.15244-493.39132	&443      & 426    	 & 101    	\\
555.21378-555.35725	&bad      & 144    	 & 133    	\\
751.39102-751.62263 & --      & 444      & 443      \\
779.54398-779.62327 & --      & 101      & 99       \\
779.38032-779.68534 & --     &  246      & 242 \\
\hline                                   
\end{tabular}
{\it Note}. The last line corresponds to observations at AO SU.
\end{table}


\subsection{Photometric observations at the Astronomical Observatory of the Siena University}

At AO SU we used a 0.30-m f/5.6 Maksutov-Cassegrain type telescope atop a Comec 10~micron GM2000-QCI equatorial mount.
Image acquisition was provided by a Sbig STL-6303 CCD camera (with a 3072~$\times$~2048 pixels of 1.15~arcsec size) and an Optec TCF-S precision focuser with temperature compensation.
The images were acquired alternating the Johnson-Cousins photometric system $V$ and $I$ filters with exposure times of 20~seconds for both filters and defocusing the telescope to increase the signal to noise ratio. HD\,27130 was observed on the night of the $22^{\rm nd}$ of October, 2019, when a secondary eclipse of HD\,27130 was expected. 

\subsection{Photometric observations with MASCARA}

In addition to the BVI photometry we also use observations obtained by the MASCARA survey \citep{Talens2017}. The MASCARA survey utilizes two stations, located in the northern and southern hemisphere, to search for transiting exoplanets. For this purpose it obtains continuous, high-cadence observations of the brightest stars in the sky (4 < V < 8.4), and such observations are also very well suited for use in variable star studies \citep{Mellon2019}. HD\,27130 passes over the East, West and Zenith pointing cameras of the northern MASCARA station and we use data obtained from February 2015 to August 2018. 
MASCARA uses no filter in its optics, as such the bandpass is determined by the transmission of the windows and lenses and the response of the CCDs. No laboratory measurement of this combined response was carried out, however investigation of the measured relative MASCARA magnitudes reveals no deviations from $V$-band magnitudes. The quantum efficiency as a function of wavelength for the CCD can be seen in fig.~4 of \citet{Talens2017}.
The raw MASCARA photometry was initially processed with the primary and secondary calibration steps of the MASCARA calibration pipeline \citep{Talens2018}, removing common mode systematics present in all light curves and residual systematics in individual light curves, respectively. In order to better disentangle the eclipse signal and the residual systematics we obtained a running mean of the fully calibrated phase-folded light curve using a period of 5.6092159 days and window size of 15 minutes. The running mean was then subtracted from the post-primary calibration light curve before re-running the secondary calibration step, and this procedure was iterated until convergence was reached. 

\subsection{Spectroscopic observations at the Mol{\. e}tai Astronomical Observatory}

Spectral observations of HD\,27130 were carried out on the f/12 1.65~m Ritchey-Chretien telescope at MAO with the high-resolution Vilnius University Echelle Spectrograph (VUES, \citealt{Jurgenson2014,Jurgenson2016}).  A log of these observations is provided in Table~\ref{table:maoobs}.
The VUES is designed to observe spectra in the 4\,000~to~8\,800~\AA~wavelength range with three spectral resolution modes ($R$~=~30\,000, 45\,000, and 60\,000).  The observations of HD\,27130 were carried out using  the 60\,000 mode, however the actual resolution of this mode that is measured using Thorium-Argon (ThAr) lines is around  68\,000. The  data  were  reduced  and  calibrated  following  standard reduction procedures which included a subtraction of the bias frame, correction for flat field, extraction of orders, wavelength calibration, and a cosmic ray removal as described by \citet{Jurgenson2016}. The spectra consist of 83 spectral orders. For most spectra the ThAr lamp spectra were taken just prior and just after each exposure and 1583 ThAr lines used for the wavelength calibration.

\subsection{Spectroscopic observations with the Hertzsprung SONG telescope}

Spectroscopic data were obtained with the Hertzsprung SONG telescope \citep{Andersen2014, Grundahl2017,Frandsen2018,Fredslund2019} at the Teide observatory on Tenerife island on $9-26$ January, 2018. A log of these observations is provided in Table~\ref{table:songobs}.
Briefly, the observations were carried out with the SONG spectrograph at a resolution
of 90\,000 (1.2" slit width) and exposure times of 1800\,s.  The spectra consist of 51 spectral orders, 
spanning $4400-6800$~{\AA}. To ensure a precise wavelength calibration, we obtained one spectrum of a 
ThAr calibration lamp just prior and just after each exposure. The number of ThAr lines used
in the wavelength calibration varied between 1053 and 1102.

\begin{table}
\caption{Information on spectroscopic observations with the MAO 1.65~m telescope.}  
\label{table:maoobs}      
\centering                          
\begin{tabular}{c c c c c }        
\hline\hline                 
Midtime of exp.   & Exp. time &  Resolution   &  Altitude 	\\
(JD-2458000)   &        (s)          &              & (deg)\\

\hline 
493.33566	&	600	&	68740	&	51.28	\\
493.34305	&	600	&	68740	&	50.89	\\
493.35039	&	600	&	68740	&	50.41	\\
512.30415	&	1800	&	68280	&	49.97	\\
521.29437	&	1800	&	68113	&	48.57	\\
525.18322	&	1800	&	68115	&	50.04	\\
525.20444	&	1800	&	68115	&	51.35	\\
525.22566	&	1800	&	68115	&	51.79	\\
537.21869	&	1800	&	68344	&	51.12	\\
555.26402	&	1800	&	68220	&	39.50	\\
575.28386	&	1800	&	68054	&	25.06	\\
576.27406	&	1800	&	68049	&	26.50	\\
578.25433	&	1800	&	68370	&	29.39	\\
587.26544	&	1800	&	68014	&	22.11	\\

\hline

\hline                                   
\end{tabular}
\end{table}

\begin{table}
\caption{Information on spectroscopic observations with SONG.}  
\label{table:songobs}      
\centering                          
\begin{tabular}{c c c c c }        
\hline\hline                 
Midtime of exp.  & Slit & Exp. time &  Flux   &  Altitude 	\\
(JD-2458000)   &      &     (s)          &      (ADU)        & (deg)\\
\hline 

128.32321  &  6  &       1800.0 &         5743 &        59.00 \\
130.44565  &  6  &       1800.0 &         7634 &        72.37 \\
131.36576  &  6  &       1800.0 &         3844 &        73.48 \\
132.42151  &  6  &       1800.0 &         6536 &        76.50 \\
133.42460  &  6  &       1800.0 &         5150 &        75.34 \\
139.30799  &  6  &       1800.0 &         5209 &        63.49 \\
140.30879  &  6  &       1800.0 &         2029 &        64.55 \\
140.33224  &  6  &       1800.0 &         2212 &        71.19 \\
141.30902  &  6  &       1800.0 &         1516 &        65.41 \\
141.38665  &  6  &       1800.0 &         4756 &        78.02 \\
143.35550  &  6  &       1800.0 &         1541 &        77.78 \\
144.33251  &  6  &       1800.0 &         4204 &        74.01 \\
144.55573  &  6  &       1800.0 &         2174 &        26.77 \\
\hline                                   
\end{tabular}
\end{table}

\section{Data reduction and analysis}
\label{data reduction} 
\begin{figure}
   \centering
    \includegraphics[width=\hsize]{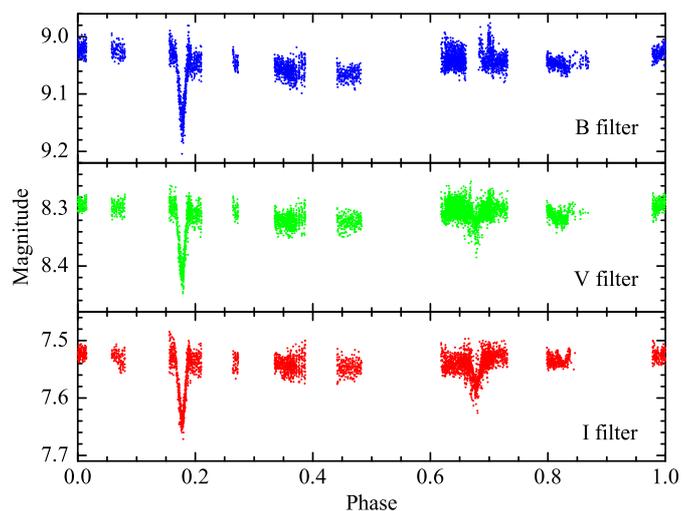}
      \caption{Phased $BVI$ light curves of HD\,27130 
      calculated with a period of 5.6092159~days.}
              
         \label{phaseMol}
   \end{figure}

\begin{figure}
   \centering
    \includegraphics[width=\hsize]{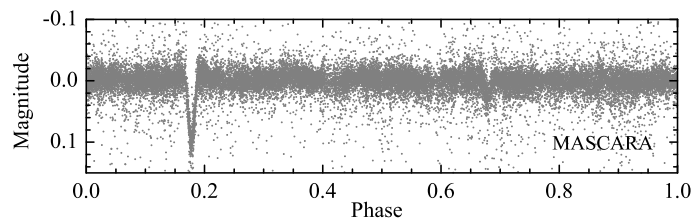}
      \caption{Phased MASCARA light curve of HD\,27130 
      calculated with a period of 5.6092159~days.}
         \label{phaseMas}
   \end{figure}

Phased light curves of the photometric observations are presented in Fig.~\ref{phaseMol} and Fig.~\ref{phaseMas}.

The photometric images observed at MAO were processed with a Muniwin program from the software package C-Munipack\footnote{http://c-munipack.sourceforge.net/}  (\citealt{Muniwin14}). The Muniwin program is built on the  basis of the software package DAOPHOT for stellar photometry in crowded stellar fields (\citealt{Daophot87}) and is designed for time series differential aperture photometry and search for variable stars. We used the {\it Advanced} image calibration procedure, to perform the bias and dark frame subtraction, and flat-field correction. For this purpose we used more than 10 images of the bias, dark and sky flat fields in each filter for the CCD image calibration of each night observations.

We used the Muniwin program to determine the instrumental magnitudes of all detected stars in the field using an aperture of 8~arcsec 
and selected among them 
one comparison star 
which was the brightest possible comparison star in the field 
and which did not show any signal indicating variability. 


For the further analysis, we calculated differential magnitudes of the target using the selected comparison star. 

Photometric analysis of images observed at AO SU was done with the MaxIm DL\footnote{http://diffractionlimited.com/product/maxim-dl/} software 
using an aperture of 14~arcsec for the $V$ filtered images and of 18~arcsec for the $I$  filtered images. Differential photometry was performed using a photometric sequence from the AAVSO Variable Star Charts service\footnote{https://www.aavso.org/variable-star-charts/}. 

We calculated apparent magnitudes of HD\,27130 in the Johnson $B$ and $V$ filters observed at MAO using HD\,285678 as a reference star. We transformed the $B_T$ and $V_T$ magnitudes of the reference star from the Tycho-2 catalogue \citep{Tycho-2-2000} to Johnson $B=11.45$ and $V=10.64$ magnitudes using equations in Appendix C of \citet{Mamajek2002}, accounting for the sign error as corrected in \citet{Mamajek2006}. There was no suitable reference star with known magnitude in $I$ filter. The observed light curve of HD\,27130 in the $V$ filter at AO~SU was fitted to the MAO observations. We could combine light curves observed at two different observatories since both observatories have observed HD\,27130 at the same phase. The apparent magnitude in the $I$ filter was calculated using the AO~SU observations, as its field of view is larger and we could use another reference star HD\,27110 recommended by AAVSO. Its value of $I$ magnitude, which is 7.852~mag, was taken from The Amateur Sky Survey TASS \citep{TASS1997}. Then the MAO observed light curve of HD\,27130 in the $I$ filter was fitted to the AO~SU observations at certain phases of the light curve.
Since the reference stars were chosen to optimize relative photometry rather than absolute, we do not use our measurements to derive precise absolute magnitudes of HD\,27130. The light curve analysis we perform later is invariant to the photometric zero-point.



\subsection{Radial velocity measurements}
For measuring radial velocities (RVs) of the binary components at each epoch, and to separate their spectra, we
used a spectral separation code based on the description of
\citet{Gonzalez2006} combined with the broadening
function (BF) formalism by \citet{Rucinski1999, Rucinski2002}.
This method works best when the observations sample
a wide range in radial velocity as evenly as possible. Calculations were done order by order. For each epoch the final radial velocity
was taken as the mean of the results from each order and the RMS scatter across orders divided by the square root of number of orders was taken as
a first estimate of the uncertainty. Later, when we fitted the
binary solution, we found that the first estimate of the RV uncertainties of the primary component were significantly smaller than the
RMS of the O–C of the best fit to the RVs. Our first solutions had O–C errors of up to 0.5~km\,s$^{-1}$ in
a systematic pattern identical for the primary and secondary
components. This indicated that the radial velocity errors were
dominated by radial velocity zero-point offsets between epochs.
Since (almost all) our spectra are calibrated with ThAr spectra taken immediately after the stellar spectra, these offsets cannot be attributed to wavelength calibration errors. 

As discussed by \citet{Griffin12}, HD\,27130 is a hierarchical triple system with a very faint tertiary star that shifts the system velocity of the primary-secondary orbit periodically with a semiamplitude of about 2.3~km\,s$^{-1}$ and a period of about 3079 days. Our observations with VUES were spread randomly in time over almost 100 days (see Table \ref{table:maoobs}) and were therefore significantly affected by this. The SONG spectra, by contrast, were all observed within just 17 days (see Table \ref{table:songobs}) and should therefore not be significantly affected. This lead us to continue with the radial velocity measurements only for the SONG spectra, given that we did not have spectra covering a long enough period to model the presence of the third component.
We did first attempt to use radial velocities from the VUES spectra by correcting for the motion of the inner pair's center of mass in the outer orbit at each epoch using the elements from \citet{Griffin12}. However, we could not reach a precision level on the orbit comparable to that with the SONG spectra. This is likely due to a combination of several issues, of which we mention some significant ones; 
1) Five VUES epochs happened to be at times with small RV differences between components, resulting in overlapping spectral lines, and thus potential systematic uncertainties on the radial velocities. 2) Although we had the elements of \citet{Griffin12} to correct for the outer orbit, we found that our measurements were not on the same RV zero-point for the SONG spectra, and even when accounting for that, we still found an additional mean offset in systemic velocity between SONG and VUES spectra of more than 1 km\,s$^{-1}$. Given a remaining larger-than-expected RV O-C scatter among the VUES spectra, it was not clear how to best subtract this additional RV offset. 3) We therefore attempted to force an RV offset so that at each VUES epoch the RV of the primary matches the orbital solution, 
and use the so obtained corrected RVs for the secondary to improve on the solution. However, this also did not work well. The secondary component RVs still had O-C scatter of the order of 1 km\,s$^{-1}$, much larger than for the SONG spectra. A significant cause for this is the lower resolution of the VUES spectra compared to SONG, meaning that more systematic noise will be smoothed into the BF peak.
Future simultaneous observations with the two instruments SONG and VUES will help improve the understanding of these issues.

Assuming that the radial velocities from SONG spectra were not significantly affected by the third component, another reason had to be found for their large O–C errors. They occurred because the
starlight cannot by exactly centred across the SONG spectrograph slit width. This shifts the spectrum in wavelength relative to a star perfectly centred across the slit, thereby mimicking a radial velocity shift. 
To correct for this shift, we used the following procedure originally developed for the analysis of \citet{Brogaard2011}: We identified a spectral order with very strong telluric absorption lines ($\lambda = 6282-6311$~\AA). In the adjacent order we measured the combined BF profile of both components using a template spectrum. Tests show that the best reproduction of the combined spectrum from the BF is obtained when the template spectrum matches the dominant contributor. Therefore, we used a template from \citet{Coelho2005} with parameters matching the primary component. The BF profile obtained in this way is nearly identical for close-by orders. We then convolved our synthetic stellar template spectrum, covering the
wavelength region with telluric absorption lines, with the obtained BF, to produce the expected stellar spectrum in that order. This spectrum was then multiplied with a synthetic telluric spectrum \citep{Bertaux2014} which was broadened to the spectrograph resolution, shifted
in radial velocity from $-1$ to +1~km\,s$^{-1}$ in steps of 10~m\,s$^{-1}$, and
multiplied with factors between 0.5 and 1.5 in steps of 0.1. The
resulting set of artificial spectra of stellar+telluric lines were all
cross-correlated with the observed spectra, and the one giving
the highest cross correlation determined the shift of the telluric
lines. This shift corresponds to the radial velocity zero-point correction, since the telluric absorption lines are always at zero radial velocity
(except for small shifts due to winds in the Earth’s atmosphere),
and the observed shift can only be caused by imperfect slit centring.
Applying the zero-point radial velocity shifts reduced the errors on the orbital parameters very significantly from O–C RMS of 300~m\,s$^{-1}$ to about 100~m\,s$^{-1}$. The uncertainties were however still underestimated. Therefore, a RV zero-point uncertainty of 100~m\,s$^{-1}$ was added in quadrature to the RV
uncertainties in order for the binary analysis to yield a reduced $\chi^2$
close to 1 for the RV of both components. The final radial velocity estimates are given in Table \ref{table:RV}.

\begin{table}
\centering
\caption{Radial velocity measurements of HD\,27130}
\label{table:RV}      
    \begin{tabular}{lrr}
\hline
\hline
BJD-TDB & RV primary & RV secondary \\
 & $({\rm km}\cdot {\rm s}^{-1})$ & $({\rm km}\cdot {\rm s}^{-1})$ \\
\hline
58128.32814 & $ 3.15\pm0.11$ & $88.50\pm0.16 $\\
58130.45043 & $98.85\pm0.11$ & $-43.47\pm0.12$\\
58131.37047 & $60.62\pm0.11$ & $ 9.41\pm0.13$\\
58132.42615 & $-5.37\pm0.11$ & $100.26\pm0.12$\\
58133.42916 &$-17.43\pm0.11$ & $116.92\pm0.12$\\
58139.31208 & $-8.25\pm0.11$ & $ 103.97\pm0.12$\\ 
58140.31280 & $52.79\pm0.11$ & $ 20.17\pm0.13$\\
58140.33625 & $54.20\pm0.11$ & $  17.77\pm0.13$\\
58141.31295 & $98.25\pm0.11$ & $-42.65\pm0.13$\\
58141.39057 & $99.39\pm0.11$ & $-43.99\pm0.13$\\
58143.35926 & $9.85\pm0.11$ &  $79.09\pm0.13$\\
58144.33618 & $-21.48\pm0.11$ & $122.88\pm0.12$\\
58144.55938 & $-19.20\pm0.11$ & $119.79\pm0.13$\\
\hline
    \end{tabular} 
\end{table}

\subsection{Light ratio}
\label{sec:LR}
The light ratio between the components is an important parameter for both the spectral separation before the spectroscopic analysis, and for the binary analysis of systems without a total eclipse, as is the case for HD\,27130.

We estimated the light ratio from an iterative procedure. First, we identified HD\,27130 in the Gaia colour-magnitude diagram (CMD) of the Hyades by \citet{Lodieu2019}. We then fitted the shape of the main sequence of the Hyades with a low-order polynomial. The light ratio in the $G$-band was estimated by requiring that both components are located on this main sequence while reproducing the system photometry when combined. The colours of the components were then used to estimate the $T_{\rm eff}$ values of the components; Guess values for $T_{\rm eff}$ were adjusted until bolometric corrections from the calibration of \citet{Casagrande2018} reproduced the component colours. These $T_{\rm eff}$ estimates were used to select template spectra from the library of \citet{Coelho2005} to fit the spectroscopic broadening functions of the separated spectra with a rotational profile. For this fit, we assumed a fixed linear limb darkening coefficient of 0.6. The ratio of the areas under the rotational profiles of the binary components corresponds to the luminosity ratio of the stars assuming that the effective temperatures are correct. The luminosity ratio is wavelength dependent, so we weighted results from the individual spectral orders in order for the averaged value to correspond to the $V$-band. This was then extended to the Gaia $G$-band by the use of Planck functions and the $T_{\rm eff}$ estimates. With the new $G$-band estimate of the luminosity ratio, the next iteration begins. In this iteration the light ratio is fixed in the CMD and both components cannot be forced to be on the main sequence. We required the primary star to be on the sequence, and allowed for the secondary to deviate. It came out a bit redder than the main sequence, which is perhaps a sign of the very faint third component. 
In few iterations, we obtained ($L_2/L_1)_V = 0.127\pm0.015$, $T_{\rm p}$ = 5660 K, and $T_{\rm s}$ = 4235 K.

\subsection{Spectral analysis}

\label{sec:specanal}

The separated spectra of the binary components resulting from the spectral separation procedure were adjusted according to the light ratio based on the Gaia CMD analysis and BF area ratios (cf. Sect.~\ref{sec:LR}) to recover the true depths of the spectral lines. This was done to remove the continuum contribution from the other component. Specifically, for the primary star, this was achieved by multiplying the separated spectrum by a factor of $\frac{L_{\rm 1} + L_{\rm 2}}{L_{\rm 1}}$ followed by a subtraction of $\frac{L_{\rm 2}}{L_{\rm 1}}$. Here, $L_{\rm 1}$ and $L_{\rm 2}$ refer to the luminosities of the primary and secondary star, respectively. We used the light ratio from the $G$-band CMD solution (cf. Sect.~\ref{sec:LR}) scaled linearly to the $V$-band. After this procedure, we estimated the S/N of the spectra as the RMS scatter in narrow wavelength-regions that were visually inspected and judged to be without spectral lines. In this way, the S/N of the separated SONG spectra were found to be roughly 150 for the primary component and 30 for the secondary at 5000~\AA, increasing to 250 and 55 at 6000~\AA. The corresponding numbers for the separated VUES spectra are 90 and 25 at 5000~\AA, and 200 and 45 at 6000~\AA.

We then performed classical equivalent width spectral analysis. This was done using two different methods, and for the SONG and the VUES spectra separately.
The first setup was as described in \citet{Slumstrup2019} with log$g$ fixed to the values from the binary solution (cf. Sect.~\ref{sec:binary}). This yielded $T_{\rm eff}$ = $5750\pm107$ K and [Fe/H]=$+0.14\pm0.02$ from the SONG spectrum of the primary star, and $T_{\rm eff}$ = $4630\pm202$ K and [Fe/H]=$+0.05\pm0.03$ from the SONG spectrum of the secondary star. The micro turbulence for the secondary was unrealistically high when unconstrained ($2.30\pm0.62$). If the micro turbulence for the secondary was fixed according to e.g. the calibration of \citet{Bruntt2010}, [Fe/H] came out higher, and in agreement with that of the primary.
Repeating the analysis for the VUES spectra gave $T_{\rm eff}$ = $5650\pm99$ K and [Fe/H]=$+0.03\pm0.02$ for the primary star, while robust results were not obtained for the secondary.

In the second setup, log\,$g$ was not fixed, but results converged naturally to the expected values.  For the spectra from SONG (VUES), the analysis made use of 161 (164) $\rm Fe\,{\sc I}$ lines and 21 (14) $\rm Fe\,{\sc II}$ lines for the primary star, and 64 (87) $\rm Fe\,{\sc I}$ lines and 2 (2) $\rm Fe\,{\sc II}$ lines for the secondary star. This yielded $T_{\rm eff} = 5614\pm40$ $(5571\pm48)$~K and [Fe/H]=$+0.00\pm0.12$ ($-0.07\pm0.10$) for the primary star and $T_{\rm eff} = 4484\pm51$ $(4534\pm63)$~K and [Fe/H]=$-0.07\pm0.11$ ($-0.05\pm0.10$) for the secondary star, using the spectra from SONG and VUES (in the brackets), respectively.

As mentioned previously, we used the Gaia CMD to establish a first-estimate of the luminosity ratio of HD\,27130. The Gaia colours obtained from this exercise resulted in the first estimate $T_{\rm eff}$ values of 5645~K and 4420~K for the primary and secondary, respectively. For the second iteration, with the light ratio constrained from the relative spectral line strengths, the $T_{\rm eff}$ of the primary changed slightly to 5660~K, while that of the secondary changed more significantly. Enforcing the spectroscopic constraint on the luminosity ratio, the secondary is no longer on the main-sequence if the colours are to be matched. In this case, the secondary $T_{\rm eff}$ becomes 4235~K, while it is 4310~K if we just choose the point on the main sequence giving the correct light ratio in Gaia $G$-band, but not for the $G_{RP}$-band.  

From all the different spectroscopic $T_{\rm eff}$ estimates and the corresponding photometric measurements, we adopt $5650\pm50$~K and $4300\pm100$~K as our best estimates of the effective temperatures. For the secondary this $T_{\rm eff}$ is somewhat lower than all our spectroscopic estimates above. However, the S/N is very low for the spectra of the secondary star, and the procedure to re-establish the true line-depths of the faint component of a spectroscopic binary is very sensitive to uncertainties in the adopted light ratio. Therefore, we put more reliance on the photometric measurements. Our after the fact comparison using the Gaia parallax in Sect.\ref{sec:dEBres} supports this choice, with agreement within 40~K.

The [Fe/H] values that we obtained are generally below the current best estimates of the metallicity of the Hyades, e.g. +0.15 \citep{Arentoft2019}. We suspect that this is due to systematic effects arising from the spectral separation and light ratio correction procedures on these relatively low S/N spectra, especially for the secondary star.

\section{Eclipsing binary analysis}

\label{sec:binary}

To determine model independent stellar parameters we used the JKTEBOP\footnote{https://www.astro.keele.ac.uk/~jkt/codes/jktebop.html} eclipsing binary code \citep{Southworth2004} which is based on the EBOP program developed by P. Etzel \citep{Etzel1981,Popper1981}. We made use of non-linear limb darkening \citep{Southworth2007}, and simultaneous fitting of the light curve and the measured radial velocities \citep{Southworth2013}. 

As seen in Fig.~\ref{phaseMol}, the light curves of HD\,27130 display small variations outside the eclipses, likely due to spot activity. Consequently, the magnitude level is different just before and after the primary eclipse compared to just before and after the secondary eclipse. To minimise this effect in the binary analysis, we kept only the parts of the light curves very close to or in eclipse. We adjusted the light curves on individual night basis such that the mean magnitude of observations taken outside eclipse aligned. We did not use the $B$-band light curve because we did not succeed in obtaining coverage of the secondary eclipse. 

For the other three filters separately, we fitted for the following parameters: orbital period $P$, time of first primary eclipse $T_0$, central surface brightness ratio $J$, sum of the relative radii $r_{\rm 1}+r_{\rm 2}$, ratio of the radii $k=\frac{r_{\rm 2}}{r_{\rm 1}}$, orbit inclination $i$, $e$cos$\omega$, $e$sin$\omega$, semi-amplitudes of the components $K_{\rm 1}$ and $K_{\rm 2}$ and system velocity of the components $\gamma_{\rm 1}$ and $\gamma_{\rm 2}$. We allowed for two system velocities because the components and their analysis could be affected differently by gravitational redshift and convective blueshift \citep{Gray2009} effects. As it turns out, they appear not to be.

We used a quadratic limb darkening law with coefficients calculated using JKTLD \citep{Southworth2015} with tabulations for the $V$ and $I$ bandpasses by \citet{Claret2000}. We ran JKTEBOP iteratively, starting with limb darkening coefficients from first guesses based on the Gaia CMD analysis of HD\,27130 and then using $T_{\rm eff}$ values from the spectral analysis. New limb darkening coefficients were then calculated with JKTLD using these $T_{\rm eff}$ and log$g$ values from the solution in the next JKTEBOP iteration.

Gravity darkening coefficients were taken from \citet{Claret2011} though even very large changes to these numbers had negligible effects as expected for nearly spherical stars. The same is true for reflection effects, which were set to zero due to our pre-analysis adjustment of the out-of-eclipse magnitude level. 

As is usually the case for eclipsing binary stars without a total eclipse, $k$, the ratio of the radii was poorly constrained by the light curve. We ran tests with $k$ fixed to different values and found that a very large range of $k$ values produced almost equally good solutions. Therefore, without an external constraint on $k$, the derived stellar parameters are too uncertain to be useful. This suggests that \citet{Schiller87} were too optimistic when estimating uncertainties in their light curve analysis. To circumvent this problem, we added the light ratio $(L_2/L_1)_V=0.127\pm0.015$ as an external constraint in JKTEBOP. This light ratio was derived using the separated spectra and the Gaia CMD (cf. Sect~\ref{sec:LR}). Using Planck functions we determined and employed also the corresponding light ratio for the MASCARA filter, which is very close to the $V$-band, since these filters have nearly the same effective wavelength despite the MASCARA filter being much broader. For the $I_C$-band we proceeded differently, since $I$-filter transmissions are not so similar. We derived the $k$ values for the other filters first and then fixed the $k$ value in the $I_C$-band solution to the weighted mean from the other two filters.

\begin{figure}
   \centering
    \includegraphics[width=\hsize]{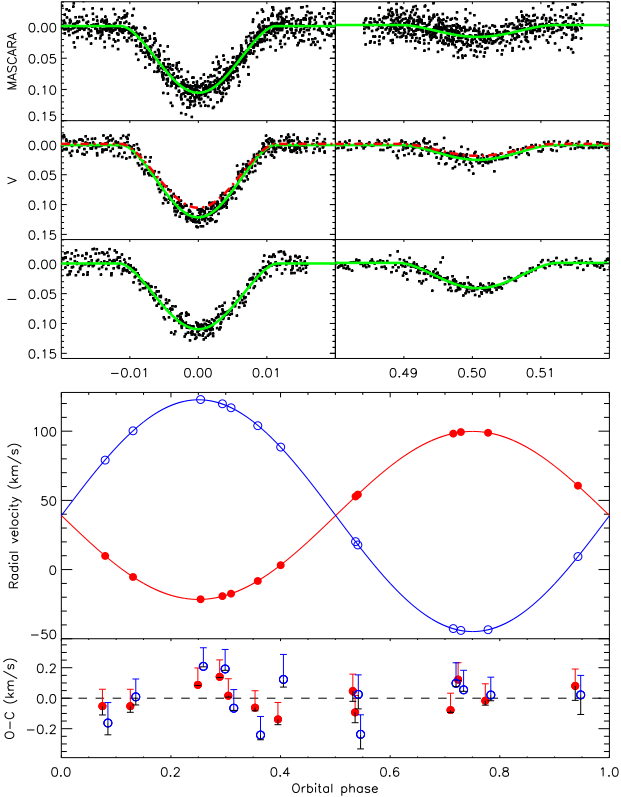}
          \caption{Light curves and radial velocity measurements of HD27130 compared to eclipsing binary models.
          Upper panels: Phased light curves showing relative magnitude in MASCARA, $V$ and $I$ filters at phases at or close to eclipse. Green lines are the best model for the given filter. In the $V$-band panel, the red dashed line is the best model using the MASCARA light curve.
          Bottom panels: Radial velocity measurements of the HD\,27130 components from SONG spectra. Red indicate the primary component and blue is the secondary. In the O–C panel, the measurements of the primary and secondary component are shifted by +/- 0.005, respectively, in phase for clarity. Note that the upwards-pointing errorbar gives the measurement error, while the downwards-pointing errorbar indicates the maximum difference between models based on different light curves.}
         \label{fig:jktebop}
   \end{figure}

The JKTEBOP solutions are given in Table~\ref{table:EBdata} and compared to the observed light curves and measured radial velocities in Fig.~\ref{fig:jktebop}. To estimate uncertainties we used the residual-permutation uncertainty estimation method of JKTEBOP, which estimates parameter uncertainties while accounting for correlated noise. The $3\sigma$ ranges of these parameter distributions correspond roughly to the filter-to-filter solution differences, and we therefore adopted these as our final $1\sigma$ uncertainties. This inflation of internal errors seems justified by the comparison of $V$-band and MASCARA light curve solutions in the middle panel of Fig.~\ref{fig:jktebop}; both limb darkening coefficients and light ratios are very close to identical for these two filters. Therefore, the difference in model eclipse depth is related to a combination of measurement errors and improper rectification of the light curves, time-variable spot patterns on the stars, and potential blended light from other sources in the MASCARA light curve - neither of which is accounted for by the internal errors. A  flat mean of the three different filter solutions were used as the best parameter estimate.

\begin{table}
\caption{Binary solutions for HD\,27130}
\label{table:EBdata}      
    \begin{tabular}{lrrr}
\hline
\hline
Parameter & $V$ & MASCARA & $I_C$ \\
\hline
Constraints: & & &\\
$L_{\rm s}/L_{\rm p}$ & 0.127(15) & 0.128(15)& --  \\
$k=(r_{\rm s}/r_{\rm p})$ & -- & -- & 0.7962\tablefootmark{*} \\
\hline
$P$ ($\rm days$) & 5.609217(7) & 5.609209(5) &  5.609233(6)   \\
Inclination \emph{i} & 85.830(54)& 85.495(112) & 85.640(19)\\
$k = r_{\rm s}/r_{\rm p}$ & 0.790(27)& 0.815(47)& 0.7962\tablefootmark{*}\\
$r_{\rm s} + r_{\rm p}$ & 0.1008(7)& 0.1053(17)& 0.1043(4) \\
$J_{\rm s}/J_{\rm p}$ & 0.2374(48)& 0.2148(121)& 0.4180(52) \\
$L_{\rm s}/L_{\rm p}$ & 0.133(15) & 0.129(15)& 0.248 \\
$\sigma (\rm{mmag})$ & 8.38 & 14.88 & 9.23 \\
\hline
$K_{\rm p} ({\rm km}\cdot {\rm s}^{-1})$ & 60.705(41)& 60.702(41)& 60.702(41) \\
$K_{\rm s} ({\rm km}\cdot {\rm s}^{-1})$ & 83.754(47)& 83.750(47)& 83.751(47) \\
$\gamma_{\rm p} ({\rm km}\cdot {\rm s}^{-1})$ & 39.049(31)& 39.039(31) & 39.040(31)\\
$\gamma_{\rm s} ({\rm km}\cdot {\rm s}^{-1})$ & 39.041(36)& 39.052(36)& 39.052(36)  \\
$e$ & 0.0013 & 0.0013 & 0.0013 \\
$\omega $ & 15.47 & 1.38 & 1.98 \\
\emph{a} $\rm(R_{\odot})$ & 16.05 & 16.06 & 16.06   \\
$M_{\rm p} \rm(M_{\odot})$ & 1.0239 & 1.0251 &  1.0246        \\
$M_{\rm s} \rm(M_{\odot})$ & 0.7421 & 0.7430 &  0.7426   \\
$R_{\rm p} \rm(R_{\odot})$ & 0.9033 & 0.9317 &  0.9330   \\
$R_{\rm s} \rm(R_{\odot})$ & 0.7139 & 0.7597 &  0.7429   \\
\hline
    \end{tabular} 
        \tablefoot{
\tablefoottext{*}{The value was fixed.} 
}
\end{table}


\begin{table}
\centering
\caption{Properties of HD\,27130}
\label{table:EBparameters}      
    \begin{tabular}{lr}
\hline
\hline
IDs & HD\,27130, vB\,22, V818\,Tau  \\
        $\alpha_{\rm J2000}$ & 04 17 38.946   \\
        $\delta_{\rm J2000}$ & +16 56 52.21  \\
        $G_{\rm TOT}$ & 8.098\\
        $V_{\rm TOT}$ & 8.315\tablefootmark{1}\\
$T_{\rm p} (\rm K)$ & 5650(50)  \\
$T_{\rm s} (\rm K)$ & 4300(100)  \\
$M_{\rm p} (\rm M_{\odot})$ & 1.0245(24)       \\
$M_{\rm s} (\rm M_{\odot})$ & 0.7426(16)      \\
$R_{\rm p} (\rm R_{\odot})$ & 0.9226(150)      \\
$R_{\rm s} (\rm R_{\odot})$ & 0.7388(260)      \\        
$V_{\rm p}$ & 8.445(14)\\
$V_{\rm s}$ & 10.685(210)\\
$BC_{V \rm ,p}$ & $-0.080$\\
$BC_{V \rm ,s}$ & $-0.746$\\
parallax, Gaia DR2 (mas) & 21.399(69)        \\
parallax, derived, primary (mas)& 21.38\\
parallax, derived, secondary (mas)& 22.33\\

\hline
    \end{tabular} 
\tablefoot{
\tablefoottext{1}{Adopted from \citet{Kharchenko2001}} 
}
\end{table}

We did not attempt to derive a precise $V$-band system magnitude for HD\,27130 from our photometry, since it was obtained with relative photometry in mind, not absolute. Instead, We estimated the apparent magnitude of the binary components by correcting the HD\,27130 system photometry in the $V$-band by \citet{Kharchenko2001} according to our spectroscopically measured $V$-band light ratio of 0.127.   
We then calculated the luminosity of the components from their radii and $T_{\rm eff}$ and used equation (10) of \citet{Torres2010} generalised to an arbitrary filter $X$ to calculate their absolute component magnitudes:
\begin{equation}
M_X=-2.5\rm{log}\left(\frac{\it L}{{\it L}_\odot}\right)+{\it V}_\odot+31.572-\left(BC_X-BC_{V,\odot}\right)
\end{equation}

Here, $V_\odot=-26.76$ as recommended by \cite{Torres2010} and BC$_{V,\odot}=-0.068$ as obtained from the calibration of \citet{Casagrande2014}. Bolometric corrections were calculated from the calibration of \citet{Casagrande2014}.

With an assumed zero interstellar reddening and absorption, we then calculated a predicted parallax for the components from the $V$-band magnitudes:

\begin{equation}
\pi = 10^{(\frac{M_V-V-5}{5})}
\end{equation}

The predicted parallaxes are given in Table \ref{table:EBparameters} along with the Gaia DR2 parallax \citep{Gaia2018}. As seen, they are in excellent agreement; a $T_{\rm eff}$ decrease of less than 10~K would be enough to obtain exact agreement for the primary. Even for the secondary, which is more sensitive to the assumed light ratio, a $T_{\rm eff}$ increase of less than 40~K is needed for exact agreement with the Gaia DR2 parallax. Given the relatively large parallax of the Hyades, the systematic parallax offset of the order of 0.05~mas in Gaia DR2 that has been investigated quite extensively by now (e.g. \citealp[and references therein]{Brogaard2018}, \citealt{Khan2019}) is impossible to address. The distance to HD\,27130 is thus 46.5~pc as calculated from the binary parameters, very close to the 46.7~pc implied by the Gaia DR2 parallax.

\label{sec:dEBres}

The final parameters and their uncertainties for HD\,27130 are given in Table~\ref{table:EBparameters}.
The masses are in agreement within uncertainties with what was obtained from spectroscopy by \citet{Griffin12}.

Our velocity semi-amplitudes, $K_{\rm p}=60.703(41)\, {\rm km\, s}^{-1}$ and $K_{\rm s}=83.752(47)\, {\rm km\, s}^{-1}$ are much more precise, but still in reasonable agreement with those derived by \citet{Griffin12}, $K_{\rm p}=60.87(6)\, {\rm km\, s}^{-1}$ and $K_{\rm s}=84.09(49)\, {\rm km \,s}^{-1}$. 
Our mass estimates are also seen to be in good agreement with the masses given in Table~17 of \citet{Griffin12} that are just very slightly larger than ours. Given that all other mass estimates we have been able to find in the literature are even larger, our mass estimates are lower than any previous estimates. We ascribe this to the effects of the third component that was extensively discussed by \citet{Griffin12}, but not properly accounted for in other studies.

\section{Re-evaluation of the properties of $\epsilon$ Tau}
\label{sec:epstau}
In order to get a tight age constraint on the Hyades age, we need a more evolved star than the components of HD\,27130. We therefore turn our attention to the giant Hyades member $\epsilon$ Tau (HD\,28305). 
The physical properties of $\epsilon$ Tau were measured by \citet{Arentoft2019} from a combination of their own asteroseismic, interferometric, and spectroscopic measurements, coupled to photometric observations and the parallax measurement from Hipparcos data \citep{vanLeeuwen2007}.
They obtained a radius of $R=12.06\pm0.16\,R_{\odot}$ and mass of $M=2.458\pm0.072\,M_{\odot}$ using a method that empirically determines also the correction needed for the asteroseismic scaling relations. However, as mentioned by \citet{Arentoft2019}, that solution suggests that $\epsilon$ Tau is an RGB star, in conflict with timescales of the secondary clump (RC2) compared to the RGB, the CMD position, and even a neural network evaluation of the oscillation power spectrum. 
Therefore, we consider the properties obtained instead by using a model-predicted correction to the asteroseismic scaling relation assuming that $\epsilon$ Tau is in the RC2 phase of evolution. Doing that, we obtained $R=12.42\,R_{\odot}$ and mass of $M=2.607\,M_{\odot}$ for a self-consistent model-predicted correction $f_{\Delta\nu}=1.005$ from the figures of \citet{Rodrigues2017} and corresponding unpublished ones showing the correction as a function of $\nu_{\rm max}$ (priv. comm. A. Miglio, see Appendix \ref{fig:scaling}). This solution requires a parallax of $\pi=21.60$ mas for an exact match between the asteroseimic and interferometric radius, 2.56 $\sigma$ away from the Hippacos measurement of $22.24\pm0.25$ mas. However, if using the $1-\sigma$ lower boundary value for $\nu_{\rm max}$, the parallax increases to $\pi=22.03$ mas, just within the $1-\sigma$ lower boundary of the Hipparcos parallax. The values of radius and mass for this solution are $R=12.17\,R_{\odot}$ and mass of $M=2.458\,M_{\odot}$, close to those of \citet{Arentoft2019}. This solution seems to be the one that best agrees with all observations. We summarize the parameters of $\epsilon$\,Tau in Table~\ref{table:epstauparameters}.

\begin{table}
\centering
\caption{Properties of $\epsilon$\,Tau}
\label{table:epstauparameters}      
    \begin{tabular}{lr}
\hline
\hline
IDs & HD\,28305, $\epsilon$\,Tau  \\
        $\alpha_{\rm J2000}$ & 04 28 36.999   \\
        $\delta_{\rm J2000}$ & +19 10 49.54  \\
        $V$ (mag) & 3.53\\
$T_{\rm eff}(\rm K)$ & $4950\pm22$\tablefootmark{*}  \\
$BC_{V}$ (mag) & $-0.257$\\
$\nu _{\rm max}(\mu$Hz) & $56.4\pm1.1$$^*$\\
$\Delta \nu(\mu$Hz) & $5.00\pm0.01$$^*$\\
$\theta _{\rm LD}$ (mas) & $2.493\pm0.019$$^*$ \\
parallax, Gaia DR2 (mas) & $20.31\pm0.43$  \\
parallax, Hipparcos (mas)& $22.24\pm0.25$  \\
\hline
\multicolumn{2}{l}{Asteroseismic parameters constrained by exact Hipparcos $\pi$,}\\
\multicolumn{2}{l}{Arentoft et al. (2019):}\\

$M (\rm M_{\odot})$ & $2.458\pm0.072$$^*$       \\
$R (\rm R_{\odot})$ & $12.06\pm0.16$$^*$      \\
\hline
\multicolumn{2}{l}{Asteroseismic parameters constrained by Hipparcos $\pi$,}\\
\multicolumn{2}{l}{This work:}\\
$\nu _{\rm max}(\mu$Hz) & $55.3$\\
$M (\rm M_{\odot})$ & 2.458       \\
$R (\rm R_{\odot})$ & 12.17      \\
parallax, derived (mas)& 22.03\\
\hline
\multicolumn{2}{l}{Asteroseismic parameters NOT constrained by Hipparcos $\pi$:}\\
\multicolumn{2}{l}{This work:}\\
$M (\rm M_{\odot}$) & 2.607       \\
$R (\rm R_{\odot}$) & 12.42      \\
parallax, derived (mas)& 21.60\\
\hline
    \end{tabular} 
\tablefoot{  
\tablefoottext{*}{Derived by \citet{Arentoft2019}} 
}    
\end{table}

Both \citet{Schroder2020} and \citet{Gray2019} find significantly larger radii for $\epsilon$ Tau and the other three well-known Hyades giants. 
\citet{Gray2019} do so while noting that for the four Hyades giants the Hipparcos and Gaia DR2 parallaxes have "disquieting differences of about 10\%", which can also be seen in Table \ref{table:epstauparameters}. We add that even the relative star-to-star differences are inconsistent. The fact that the Gaia DR2 parallax of $\epsilon$\,Tau is $20.31\pm0.43$~mas, many $\sigma$ away from the Hipparcos value calls for reconsideration of the physical properties in the future, when Gaia parallaxes of bright stars are better estimated.

Part of the larger radii found by both \citet{Schroder2020} and \citet{Gray2019} is however not related to the parallax, but rather the use of bolometric corrections (BCs) that are quite different from the ones we adopt from \citet{Casagrande2014}. \citet{Schroder2020} adopts a BC$_{V}$ of $-0.5$~mag, mentioning also the consequences for radius and mass given a BC of $-0.4$~mag. \citet{Gray2019} do not mention the source or value of the BCs they use, but reversing their calculations of the radii under the assumption that BC$_{V,\odot}=-0.068$ \citep{Casagrande2014}, we found that they used BC$_V=-0.39$ for $\epsilon$ Tau (and $\delta$ Tau, and BC$_V=-0.36$ for the other two giants).

The value we use for $\epsilon$\,Tau, corresponding to $T_{\rm eff}=4950$~ K and log$g=2.7$ from \citep{Casagrande2014} is BC$_V=-0.257$.
By comparing the physical radius obtained from interferometry to the physical radius obtained from the spectroscopic $T_{\rm eff}$ and $V$-band magnitude for the same parallax value, the validity of the bolometric correction can be tested. Doing this, we confirmed the excellent agreement for our value of BC$_V=-0.257$ that was also established in \citet{Arentoft2019}; the two radius estimates agree within 0.2\%. Using instead the bolometric corrections suggested by \citet{Schroder2020} or \citet{Gray2019}, the radii become 12\% or 7\% larger than the interferometric radius. This disagreement remains regardless of which parallax is adopted, suggesting that the $BC_V$ of \citet{Casagrande2014} is the more accurate one.

\section{Stellar models and isochrone comparisons}
\label{sec:results}
We used Victoria models \citep{VandenBerg2014}, PARSEC models \citep{Bressan2012}, and two different MESA model grids \citep{Paxton2011,Paxton2018}.

The Victoria models assume the \citet{Asplund2009} solar abundances, scaled to different [Fe/H] values. \citet{Choi2016} find that an initial $Z = 0.015$ and $Y = 0.261$ are reduced to the Asplund $Z$ and $Y$ consistent with asteroseismology when diffusion acts for the solar age (4.57~Gyr). For the Victoria models generated here:

[Fe/H] = 0.10, $Y$ = 0.27,  $Z$ = 0.0163; 

[Fe/H] = 0.10, $Y$ = 0.30,  $Z$ = 0.0155; 

[Fe/H] = 0.15, $Y$ = 0.27,  $Z$ = 0.0182; \\
one might expect that the Hyades would have $Y$ fairly close to 0.27 if it has [Fe/H] between 0.10 and 0.15 suggested by spectroscopic investigations (e.g. \citet{Arentoft2019}. Victoria models do not include diffusion of metals, only settling of helium, but at an age of <~1~Gyr, the effects of metals diffusion will be minimal.

The PARSEC model are those of \citet{Bressan2012} with revised and calibrated surface boundary conditions in low-mass dwarfs described by \citet{Chen2014}.

We use two different versions of MESA models, which both include diffusion of helium and metals; The MIST models \citep{Dotter2016,Choi2016} that assume the \citet{Asplund2009} solar abundances and those from a specific grid of \citet{Miglio2020} that assume \citet{Grevesse1993}\ solar abundances. The latter models are as described in \citet{Rodrigues2017} except that they include diffusion.

\subsection{Model comparisons to HD\,27130}

\begin{figure}
   \centering
    \includegraphics[width=\hsize]{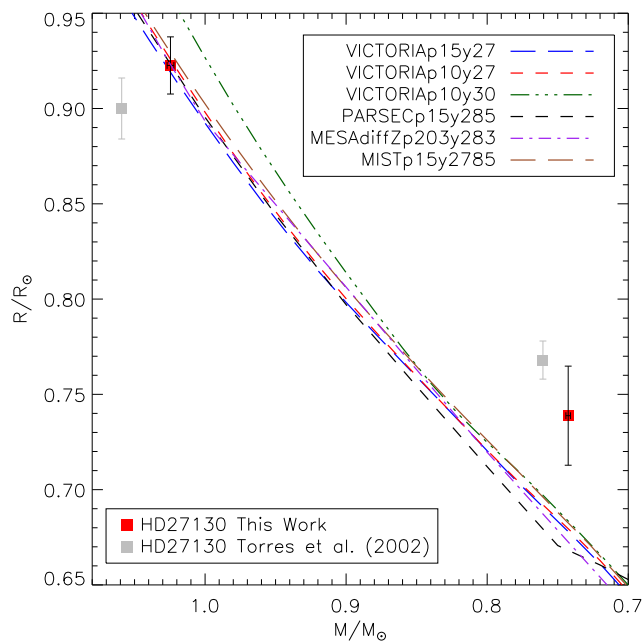}
    \includegraphics[width=\hsize]{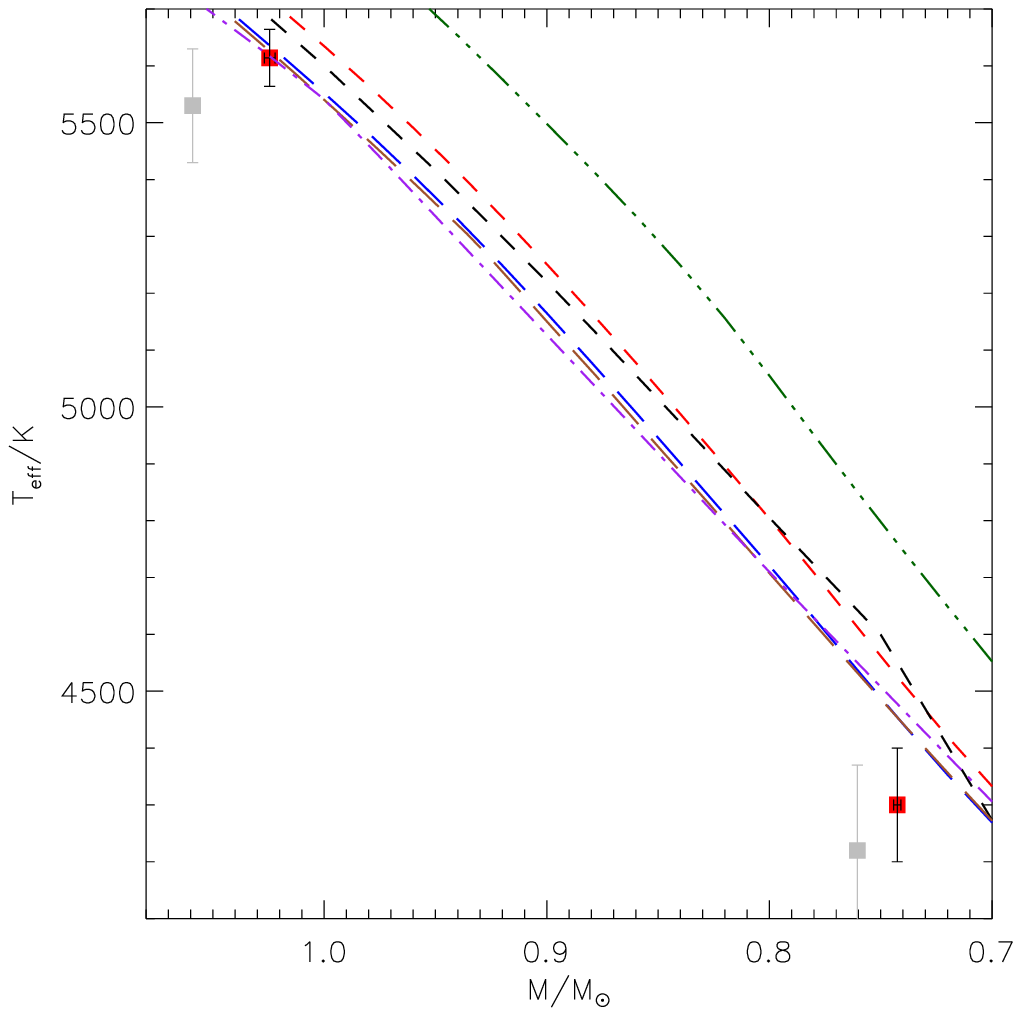}
      \caption{Mass-Radius (upper panel) and Mass-$T_{\rm eff}$ (lower panel) diagrams comparing the components of HD\,27130 to isochrones from different model sets and different compositions. Notation is such that e.g. p15y27 means [Fe/H]$=+0.15$ and $Y=0.27$. 
      The red squares are the measurement from this work, while the gray squares are the corresponding values from \citet{Torres02}.}
         \label{fig:M-R-T}
   \end{figure}

Fig.~\ref{fig:M-R-T} shows the Mass-Radius and Mass-$T_{\rm eff}$ diagrams of the HD\,27130 components compared to selected isochrones. The primary component matches most of the isochrones well; In the mass-radius diagram, all isochrones except the one assuming a very high helium content agree well within 1 $\sigma$. In the Mass-$T_{\rm eff}$ diagram, both the high helium and lower metallicity Victoria isochrones and the PARSEC isochrone are hotter than the HD\,27130 primary. The secondary is larger and cooler than all the isochrones. For comparison, we also show the measurement by \citet{Torres02}. They also found the secondary to be larger and cooler than the isochrones, but their estimates for the primary also does not match the isochrones shown.

At the mass range spanned by the components of HD\,27130, the isochrones have no significant dependence on age. Therefore, the measurements of HD\,27130 can be used to make predictions about the helium content of the Hyades by adopting a measured [Fe/H] for the cluster.

We disregard the secondary component, since it is larger and cooler than all the isochrones regardless of the composition. This is likely related to inflation caused by magnetic fields, as suggested for other binary components of similar mass in the literature (e.g. \citealt{Sandquist2016}). The PARSEC isochrones are corrected empirically to match the lower main-sequences of open and globular clusters \citep{Chen2014}. This can be seen in both panes of Fig.~\ref{fig:M-R-T} where the PARSEC isochrones bend slightly at a mass very close to the mass of the HD\,27130 secondary. However, the effect is too insignificant to make a difference , which is in accord with the findings of \citet{Chen2014} that the effects are very small at this mass. Thus, the inflated radius and decreased $T_{\rm eff}$ of the HD\,27130 secondary relative to isochrone predictions are likely caused by an increased magnetic activity relative to single low mass stars.

Comparing the position of the primary to the Victoria isochrones, we estimate $Y=0.274\pm0.017$ from the Mass-Radius diagram and $Y=0.267\pm0.006$ from the Mass-$T_{\rm eff}$ diagram at [Fe/H]=$+0.15$. $Y=0.27$ implies a helium enrichment law close to $\frac{\Delta Y}{\Delta Z}=1.2$ when calculated from a primordial value of $Y=0.248$. This is in agreement with $\frac{\Delta Y}{\Delta Z}=1.4\pm0.1$ estimated for the open cluster NGC6791 by \citet{Brogaard2012} through analysis of eclipsing members. It thus appears that the helium enrichment of the Hyades is similar to other open clusters. 

The MIST models, which also assume the \citet{Asplund2009} solar abundances, suggest very close to the same helium content, $Y=0.274$ from the Mass-Radius diagram but $Y=0.277$ from the Mass-$T_{\rm eff}$ diagram. This illustrates that different assumptions about model physics contributes at least 0.005 the uncertainty in $Y$ when determined from the Mass-Radius and Mass-$T_{\rm eff}$ diagrams.

The PARSEC models assume another solar abundance reference \citep{Caffau2011}. They suggest $Y=0.285$ from the Mass-Radius diagram but $Y=0.277$ from the Mass-$T_{\rm eff}$ diagram. A higher [Fe/H] is needed in order for these isochrones to yield a consistent value for $Y$. Therefore, although these models suggest a larger helium content, the predicted helium enrichment law will be similar, because of the corresponding larger $Z$ value.


The lower helium content of $Y=0.255$ derived by \citet{Lebreton2001} from HD27130 was likely an artifact caused mainly by the previous mass overestimate due to the unseen third component.

\begin{figure}
   \centering
    \includegraphics[width=\hsize]{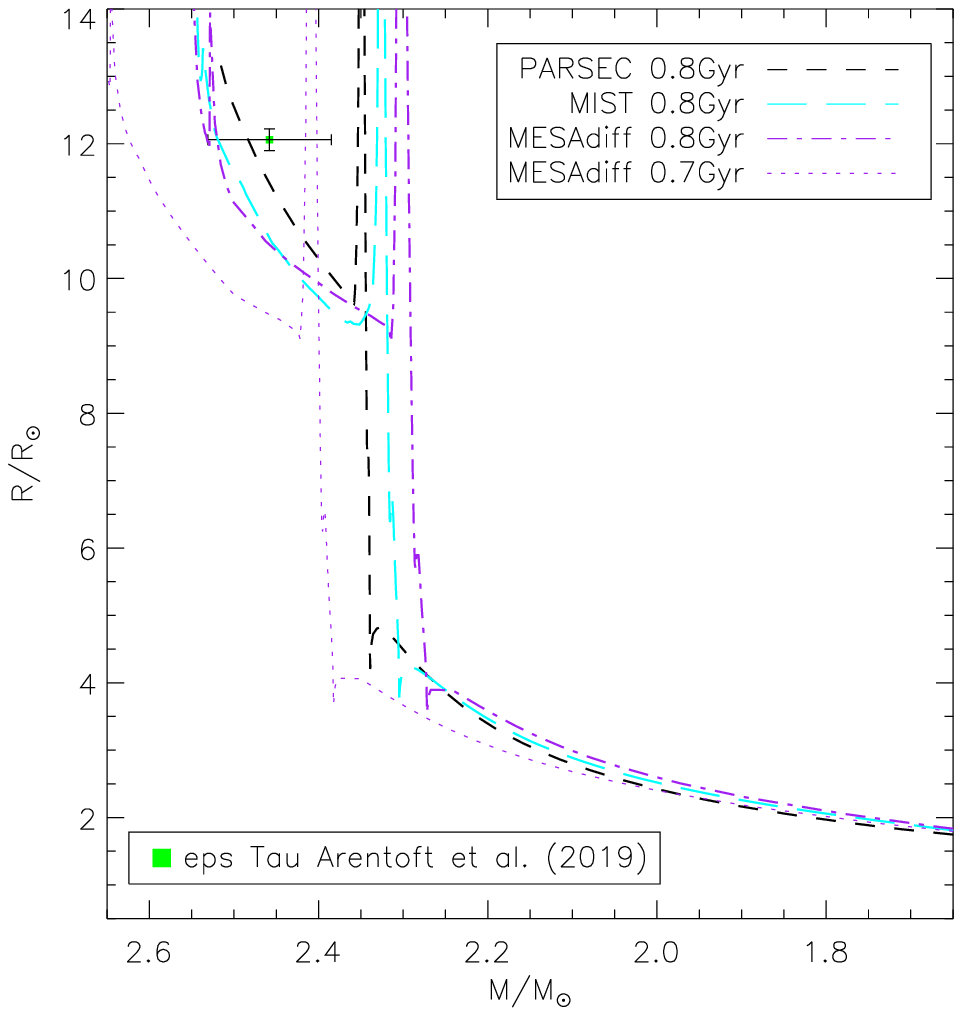}
    \includegraphics[width=\hsize]{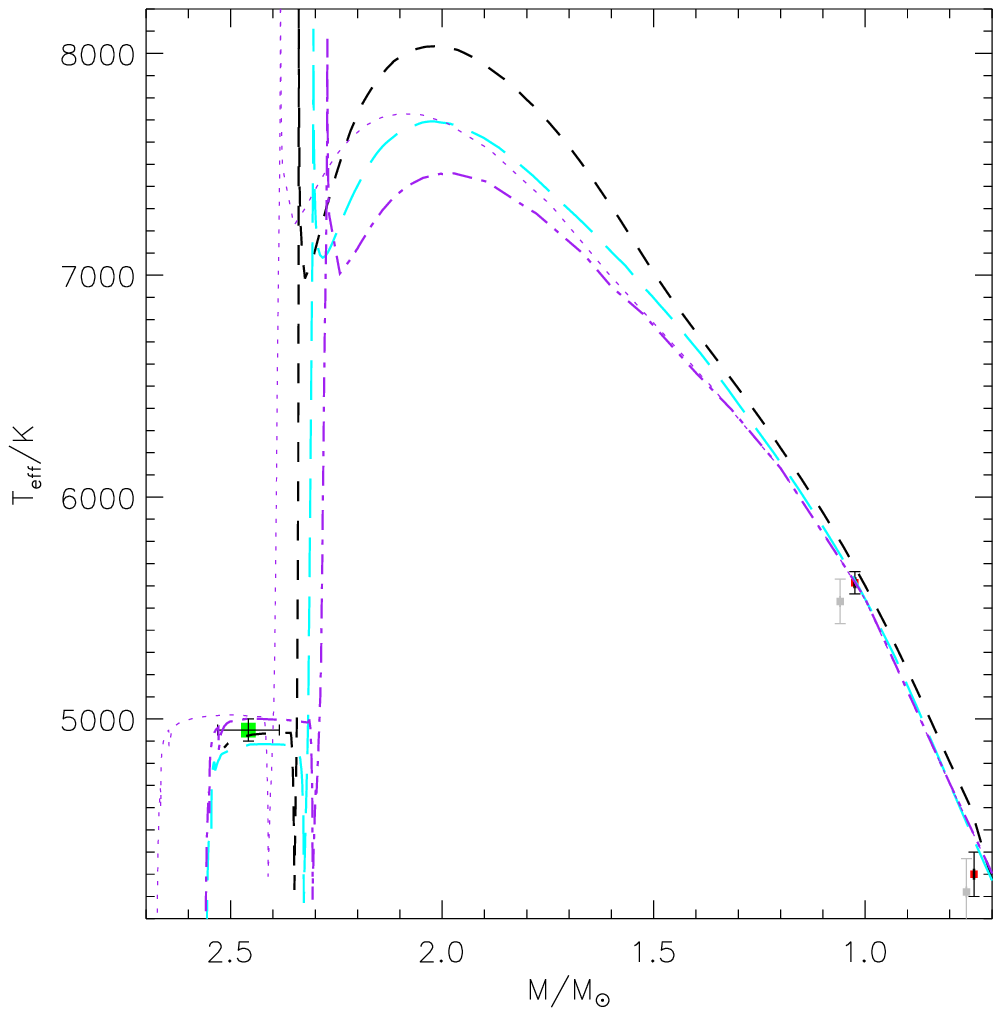}
      \caption{Mass-Radius (upper panel) and Mass-$T_{\rm eff}$ (lower panel) diagrams comparing the asteroseismic measurements of $\epsilon$ Tau to isochrones from different model sets and different compositions. The green square is the measurement of $\epsilon$ Tau from \citet{Arentoft2019}. The red squares are the measurements of HD27130 from this work, while the gray squares are the corresponding values from \citet{Torres02}.}
         \label{fig:M-R-T3}
   \end{figure}

\subsection{Model comparisons to $\epsilon$\,Tau}

With the helium content constrained by the primary component of HD\,27130, we can constrain the age of the Hyades using an evolved star. We make use of the secondary clump star member $\epsilon$ Tau with properties determined by \citet{Arentoft2019} as re-evaluated in the present paper. Since our best estimates of the mass and radius were very close to those of \citet{Arentoft2019} we use their measured values in Fig.~\ref{fig:M-R-T3}, where we compare to some of the same isochrones as in Fig.~\ref{fig:M-R-T}. As seen in the upper panel, an age close to $0.8$ Gyr is compatible with the observations if $\epsilon$ Tau is in the secondary clump phase, as the evidence strongly suggests.   

The age needed for an exact match at the measured mass of $\epsilon$ Tau is $0.90$~Gyr for the MESA isochrones and $0.83$~Gyr for the PARSEC isochrone (The Victoria models were not extended to the core He-burning phase.). The size of the $1-\sigma$ uncertainty in mass can be seen to correspond to a 0.1~Gyr $1-\sigma$ uncertainty in age. It is also worth noting that due to the mass-radius correlation in the asteroseismic estimates, the best age estimates are younger by only about 0.1~Gyr if we adopt our alternative mass and radius estimates of $\epsilon$ Tau that was not consistent with the Hipparcos distance.

The lower panel of Fig.\ref{fig:M-R-T3} compares $\epsilon$ Tau to the same models in the mass-$T_{\rm eff}$ diagram. The effective temperatures of all models are within or just outside the 1-$\sigma$ uncertainty in $T_{\rm eff}$ although different models show differences of more than 100 K for the same age. Unfortunately, the model $T_{\rm eff}$ of evolved stars is sensitive to e.g. the calibration of the mixing length parameter, adopted surface boundary conditions, the detailed abundance pattern, and convective overshooting. Therefore, it makes little sense to attempt to force a finer match within the level of the 1-$\sigma$ error-bar. While all of the many spectroscopic investigations of the Hyades giants have found their $T_{\rm eff}$ to be below 5000 K, some of the mentioned model assumptions may be responsible for the model reaching such high $T_{\rm eff}$. The two isochrones with identical assumptions except the age clearly demonstrate that the $T_{\rm eff}$ of $\epsilon$ Tau is insensitive to the assumed age. Instead, the good level of agreement show that these stellar models self-consistently predict the correct $T_{\rm eff}$ of the secondary clump ($\epsilon$ Tau) within errors if constrained to do so at the unevolved part of the main-sequence (HD\,27130).         

\subsection{Mass-luminosity and Gaia CMD analysis}

\begin{figure}
   \centering
    \includegraphics[width=\hsize]{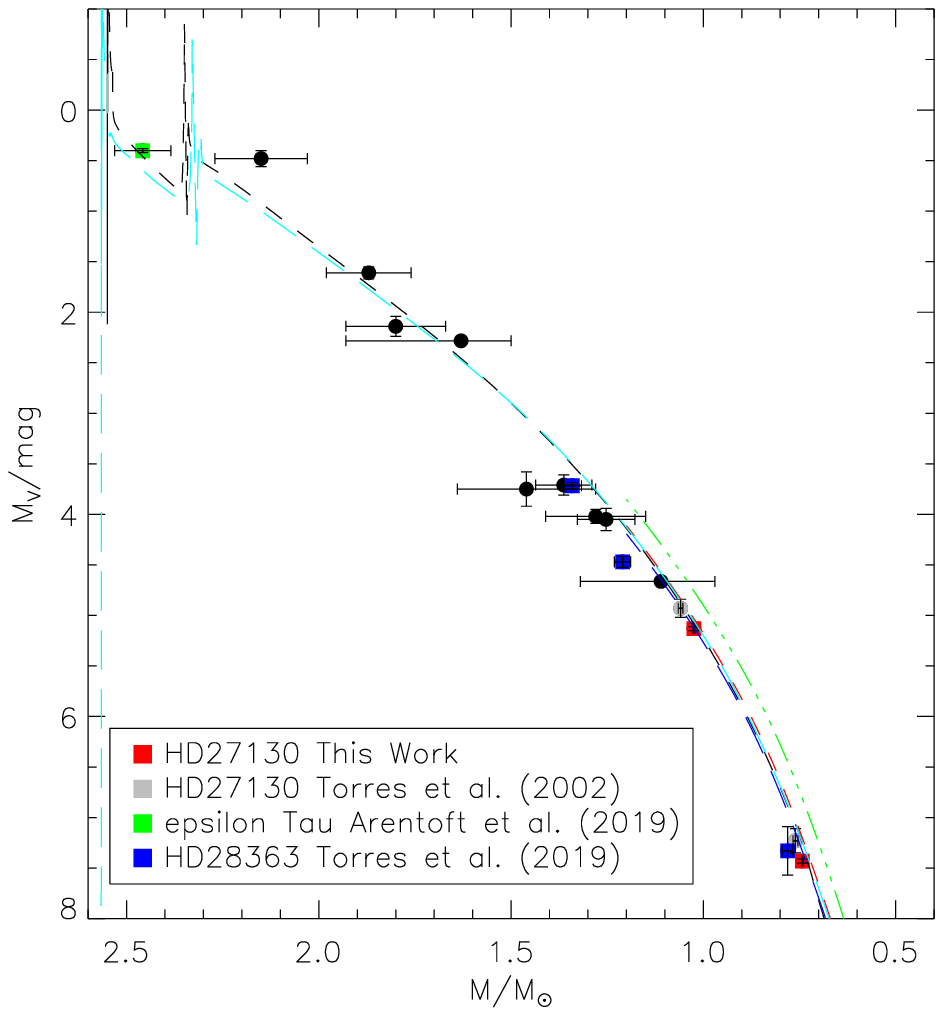}
      \caption{Mass-$M_V$ diagram comparing the components of HD\,27130 to isochrones from different model sets and different compositions. The red squares are the measurement from this work. The gray squares are the corresponding values from \citet{Torres02} according to \citet{Torres2019B} (we were unable to recover those numbers in \citealt{Torres02}). Blue squares are measurements of the triple HD\,28263 by \citet{Torres2019A}. Black circles are measurements of additional binary stars belonging to the Hyades, measured by different authors and presented by \citet{Torres2019A, Torres2019B}. The green square is the asteroseismic measurements of $\epsilon$ Tau by \citet{Arentoft2019}. Isochrones are a selected subset of those given in the legend of Fig. \ref{fig:M-R-T} Those covering the full mass range are both of age 800~Myr. The Victoria models are only calculated for low masses where the isochrone shape is not significantly affected by age.}
         \label{fig:M-L}
   \end{figure}

Fig. \ref{fig:M-L} reproduces the Mass-$M_V$ diagram of \citet{Torres2019B} with the addition of our measurements of HD27130 and the measurements of $\epsilon$ Tau by \citet{Arentoft2019}. Some of the isochrones from previous figures are also shown for comparison. Although the positions of the HD\,27130 components in this diagram have changed with our measurements relative to those of \citet{Torres02}, they have shifted along the isochrone in such a way that the same isochrone would be preferred from either set of measurements. The measurements of $\epsilon$ Tau also falls nicely along the model predictions. 

\begin{figure}
   \centering
    \includegraphics[width=\hsize]{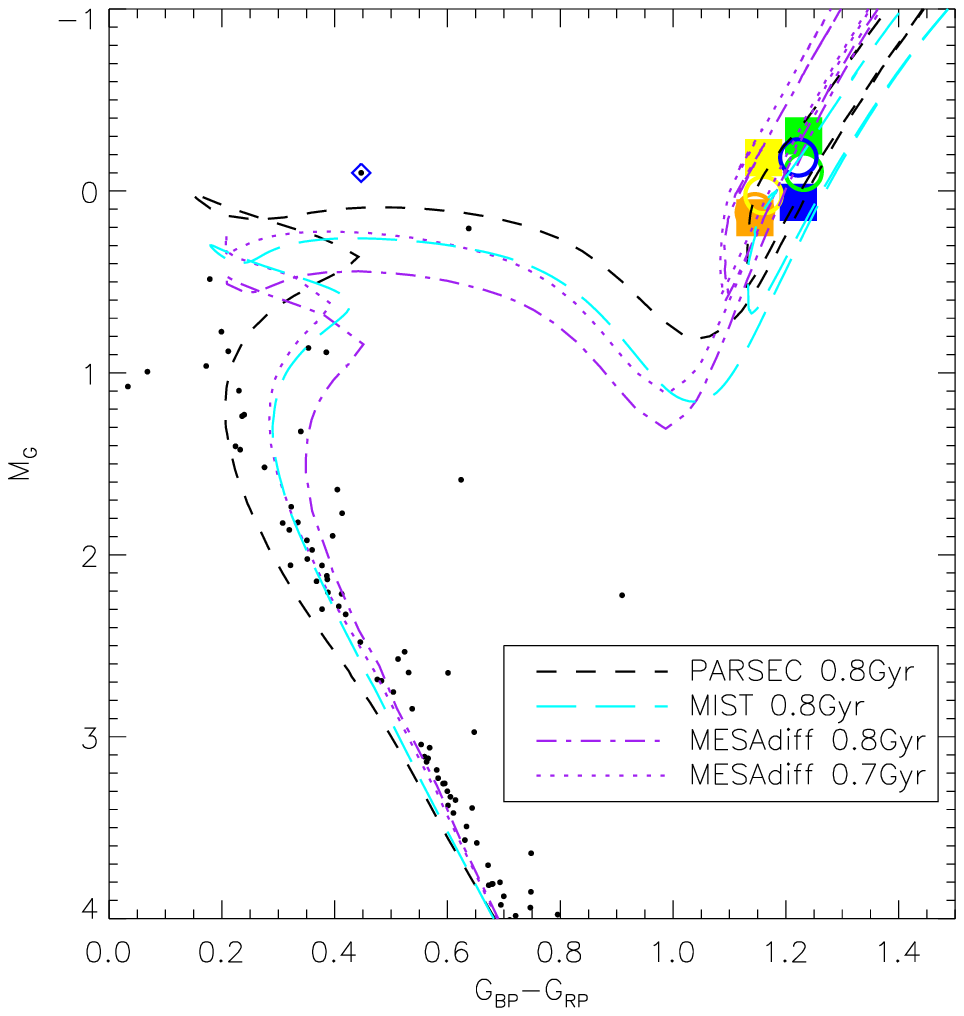}
    \includegraphics[width=\hsize]{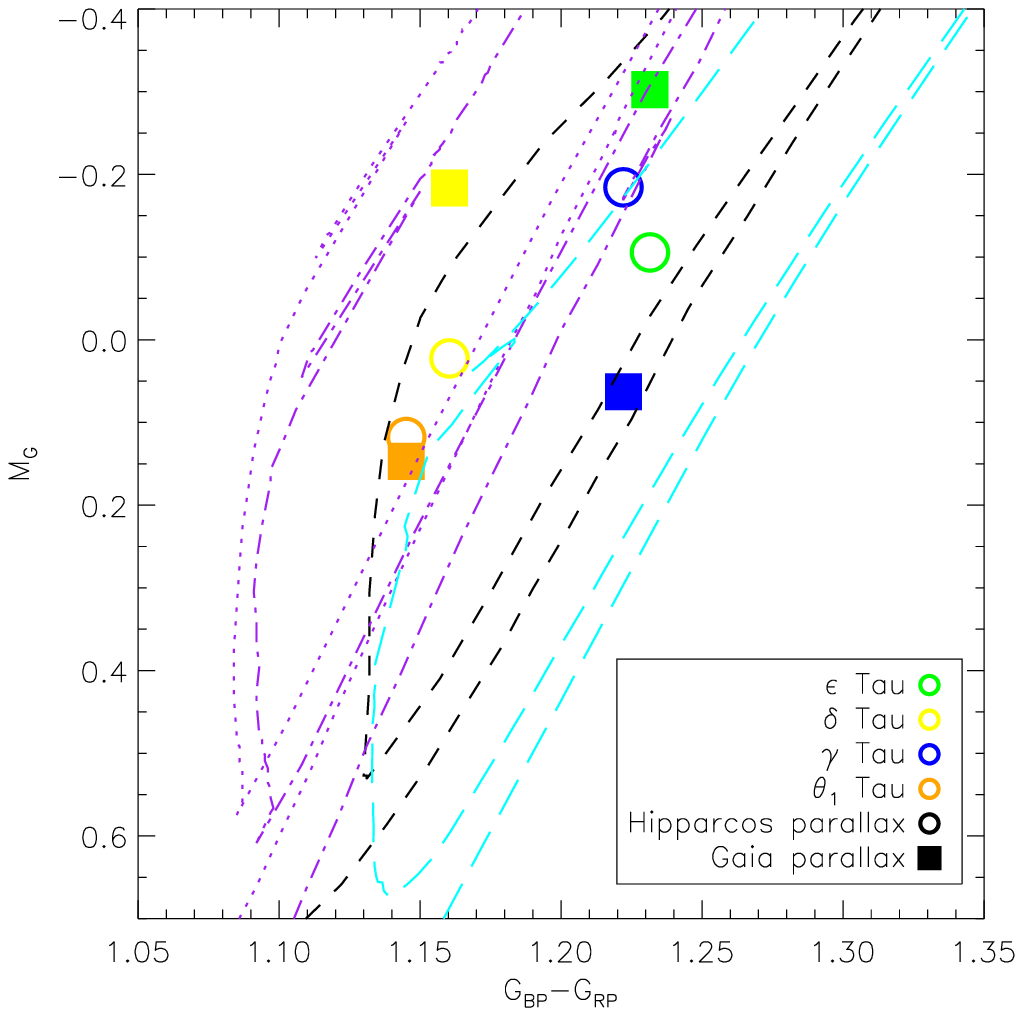}
      \caption{Upper panel: Gaia colour-magnitude diagram of the Hyades with members from \citet{Lodieu2019} compared to isochrones with the same ages as in Fig. \ref{fig:M-R-T3} and compositions as in Fig. \ref{fig:M-R-T}. 
      Lower panel: Zoom on the giant part of the same colour-magnitude diagram. The positions of the four well-known Hyades giants are marked. Circles indicate their positions when adopting the Hipparcos parallaxes. Squares mark the corresponding positions if adopting the Gaia parallaxes.}
         \label{fig:CMD}
   \end{figure}

In the upper panel of Fig. \ref{fig:CMD} we compare the same models as in Fig. \ref{fig:M-R-T3} to the Gaia CMD of selected Gaia members from \citet{Lodieu2019}. The $M_G$ magnitudes have been obtained by adjusting the $G$ magnitudes according to their parallax. The four giants have been marked individually. $\theta_{1}$\,Tau was added, since it was not selected as a member by \citet{Lodieu2019}. 

As seen, the upper main sequence and turn-off of these models are quite different even though they predict a very similar mass for $\epsilon$\,Tau. However, unfortunately, the upper main sequence close to the terminal age main sequence (TAMS) is very poorly defined, and there is at most measurements for one subgiant star. Without further investigations, it would not be well justified to demand that isochrones need to match this one star. 
The star marked with a blue diamond is $\theta_{2}$\,Tau, a known binary where two stars contribute to the combined light. Isochrones should therefore not match this point. Further investigations of the stars close to the TAMS would be needed if tight isochrone constraints are to be obtained from this Gaia CMD. However, it does appear that the PARSEC models are too blue as they fail to match the upper main-sequence stars of the Hyades, which is consistent with the isochrones being too hot compared to the HD\,27130 primary in Fig.~\ref{fig:M-R-T3}. The 0.8~Gyr MESA isochrone is somewhat too red before it bends back to the blue at the end of the core H-burning phase.

The lower panel of Fig. \ref{fig:CMD} shows only the four giants in the CMD. Squares mark their CMD positions if using the Gaia DR2 parallaxes, while circles indicate their positions if Hipparcos parallaxes are used. As seen, it makes a very significant difference if one or the other measures of parallaxes are used. On top of that, the $T_{\rm eff}$ differences among the four giants implied by the relative $G_{BP}-G_{RP}$ colours amount to a range of 218~K, while relative spectroscopic studies suggest a much smaller $T_{\rm eff}$ range of only 50~K or less \citep{Gray2019, Schroder2020}. $B-V$ colour differences suggest a range of 130~K among the giants.

This causes huge complications in trying to compare the Hyades giants to Gaia CMD observations. If, however, these four giants all belong to the secondary clump, as evolutionary timescales, their similar spectroscopic $T_{\rm eff}$ and the asteroseismic analysis of mixed modes \citep{Bedding2011,Arentoft2017,Arentoft2019} in $\epsilon$\,Tau all suggest, then the lower panel of Fig.\ref{fig:CMD} implies that the Hipparcos parallaxes are the correct ones. The Gaia DR2 parallaxes simply position the giants too far away from each other in both colour and magnitude for them to be matched by the secondary clump phase of \emph{any} isochrone. We checked that applying the Gaia saturation effect corrections in appendix B of \citet{Evans2018} did not change this conclusion significantly. Applying these corrections changes the magnitudes of all fours giant by close to identical amounts, which is not surprising, since their magnitudes are similar. The $G$-mags change by very close to +0.11~mag for all four stars, moving them closer to the isochrone predictions for the clump phase. The $G_{BP}-G_{RP}$ colours change by less than 0.01 mag for each star, but in such a way that the colour-range covered by the stars decreases by 0.016 mag corresponding to a reduction in the photometric $T_{\rm eff}$ range by 38~K to 180~K, which is still very large. Therefore, if saturation effects are a play, we suspect that random star-to-star saturation variations are significant. We have not applied saturation corrections in Fig.~\ref{fig:CMD}.

\subsection{Comparisons to other cluster age estimates.}

Our investigations above suggest that the age of the Hyades is 0.9 $\pm$ 0.1 (stat) $\pm$ 0.1 (sys)~Gyr. 
This age estimate is significantly larger than the $588\pm60$~Myr found by \citet{Schroder2020}. Part of the age difference (60~Myr) is explained by differences in the bolometric corrections as discussed above and in \citet{Schroder2020}. Thus, in light of our investigation of the bolometric corrections for the Hyades giants, the best age estimate from \citet{Schroder2020} is instead $648\pm60$~Myr. This is still younger than our age estimate.

The age of the Hyades derived in this work is also larger than the white dwarf (WD) cooling age of $640^{+67}_{-49}$~Myr found by \citet{El-Badry2018}. However, we believe that their younger age could be partly due to their assumption of solar composition for the calculation of the pre-WD evolution time, and perhaps also the assumption of solar composition in the derivation of the initial-final mass relation (IFMR). \citet{Salaris2018} derive the IFMR for the Hyades by {\it assuming} an age of 800 Myr, closer to the age that we found. This IFMR is not very different from the one by \citet{El-Badry2018} for the lower-mass WDs, but deviates for the higher-mass WDs. This indicates that perhaps the slight age discrepancy arises due to inaccuracies in the model treatment of convective core overshooting, which affects the pre-WD age, and thus the initial masses.

\section{Summary, conclusions and outlook}
\label{conclusions}

We used new observations to establish the physical properties of the components of the eclipsing Hyades member HD\,27130 as given in Table \ref{table:EBparameters}. The properties of the primary component were used to constrain the helium content, which we found to be $Y=0.27$, corresponding to a helium enrichment law close to $\frac{\Delta Y}{\Delta Z}=1.2$.

The properties of $\epsilon$\,Tau were re-analysed, finding that the mass and radius established by \citet{Arentoft2019} seem robust. A higher mass is only likely if the Hipparcos parallax of $\epsilon$\,Tau is too large by $2.56 \sigma$.   

We estimated the age of the Hyades to be $0.9\pm0.1$(stat) $\pm0.1$~(sys)~Gyr in slight tension with recent age estimates based on the cluster white dwarfs and based on the Hyades giants, but without asteroseismic constraints. 

The age precision can be much improved through asteroseismology of the other three Hyades giants in a similar fashion as was done for $\epsilon$\,Tau by \citet{Arentoft2019}. That would reduce the random error on the age, and help assessing the accuracy of the Hipparcos parallaxes of the Hyades giants. Potential future improvements to the Gaia parallax of bright stars would also help clarify the situation.
Only then the age of the Hyades can be accurately established and used to investigate convective core overshoot through its effects on the turn-off morphology, the mass of the helium burning giants, and the WD cooling sequence. We have taken the first steps down this path by providing constraints on the helium content from HD\,27130 and investigating the constraints on the mass of the giant $\epsilon$\,Tau.

\begin{acknowledgements}
We gratefully acknowledge the grant from the European Social Fund via the Lithuanian Science Council (LMTLT) grant No. 09.3.3-LMT-K-712-01-0103.\\
Observations were partially made with the Hertzsprung SONG telescope operated on the Spanish Observatorio del Teide on the island of Tenerife by the Aarhus and Copenhagen Universities and by the Instituto de Astrofísica de Canarias.\\
This work has made use of data from the European Space Agency (ESA) mission
{\it Gaia} (\url{https://www.cosmos.esa.int/gaia}), processed by the {\it Gaia}
Data Processing and Analysis Consortium (DPAC,\url{https://www.cosmos.esa.int/web/gaia/dpac/consortium}). Funding for the DPAC has been provided by national institutions, in particular the institutions
participating in the {\it Gaia} Multilateral Agreement.\\
This research has made use of the SIMBAD database,
operated at CDS, Strasbourg, France\\
Funding for the Stellar Astrophysics Centre is provided by The Danish National Research Foundation (Grant agreement no.: DNRF106).\\
AM acknowledges support from the ERC Consolidator Grant funding scheme (project ASTEROCHRONOMETRY, \url{https://www.asterochronometry.eu}, G.A. n. 772293).

\end{acknowledgements}

\bibliographystyle{aa} 
\bibliography{References-1} 

\begin{thebibliography}{89}
\expandafter\ifx\csname natexlab\endcsname\relax\def\natexlab#1{#1}\fi

\bibitem[{{Andersen} {et~al.}(2014){Andersen}, {Grundahl},
  {Christensen-Dalsgaard}, {Frandsen}, {J{\o}rgensen}, {Kjeldsen}, {Pall{\'e}},
  {Skottfelt}, {S{\o}rensen}, \& {Weiss}}]{Andersen2014}
{Andersen}, M.~F., {Grundahl}, F., {Christensen-Dalsgaard}, J., {et~al.} 2014,
  in Revista Mexicana de Astronomia y Astrofisica Conference Series, Vol.~45,
  Revista Mexicana de Astronomia y Astrofisica Conference Series, 83

\bibitem[{{Arentoft} {et~al.}(2017){Arentoft}, {Brogaard}, {Jessen-Hansen},
  {Silva Aguirre}, {Kjeldsen}, {Mosumgaard}, \& {Sand quist}}]{Arentoft2017}
{Arentoft}, T., {Brogaard}, K., {Jessen-Hansen}, J., {et~al.} 2017, \apj, 838,
  115

\bibitem[{{Arentoft} {et~al.}(2019){Arentoft}, {Grundahl}, {White},
  {Slumstrup}, {Handberg}, {Lund}, {Brogaard}, {Andersen}, {Silva Aguirre},
  {Zhang}, {Chen}, {Yan}, {Pope}, {Huber}, {Kjeldsen}, {Christensen-Dalsgaard},
  {Jessen-Hansen}, {Antoci}, {Frandsen}, {Bedding}, {Pall{\'e}}, {Garcia},
  {Deng}, {Hon}, {Stello}, \& {J{\o}rgensen}}]{Arentoft2019}
{Arentoft}, T., {Grundahl}, F., {White}, T.~R., {et~al.} 2019, \aap, 622, A190

\bibitem[{{Asplund} {et~al.}(2009){Asplund}, {Grevesse}, {Sauval}, \&
  {Scott}}]{Asplund2009}
{Asplund}, M., {Grevesse}, N., {Sauval}, A.~J., \& {Scott}, P. 2009, \araa, 47,
  481

\bibitem[{{Bedding} {et~al.}(2011){Bedding}, {Mosser}, {Huber},
  {Montalb{\'a}n}, {Beck}, {Christensen-Dalsgaard}, {Elsworth}, {Garc{\'\i}a},
  {Miglio}, {Stello}, {White}, {De Ridder}, {Hekker}, {Aerts}, {Barban},
  {Belkacem}, {Broomhall}, {Brown}, {Buzasi}, {Carrier}, {Chaplin}, {di Mauro},
  {Dupret}, {Frandsen}, {Gilliland }, {Goupil}, {Jenkins}, {Kallinger},
  {Kawaler}, {Kjeldsen}, {Mathur}, {Noels}, {Silva Aguirre}, \&
  {Ventura}}]{Bedding2011}
{Bedding}, T.~R., {Mosser}, B., {Huber}, D., {et~al.} 2011, \nat, 471, 608

\bibitem[{{Bertaux} {et~al.}(2014){Bertaux}, {Lallement}, {Ferron}, {Boonne},
  \& {Bodichon}}]{Bertaux2014}
{Bertaux}, J.~L., {Lallement}, R., {Ferron}, S., {Boonne}, C., \& {Bodichon},
  R. 2014, \aap, 564, A46

\bibitem[{{Brandt} \& {Huang}(2015)}]{Brandt15}
{Brandt}, T.~D. \& {Huang}, C.~X. 2015, \apj, 807, 58

\bibitem[{{Bressan} {et~al.}(2012){Bressan}, {Marigo}, {Girardi}, {Salasnich},
  {Dal Cero}, {Rubele}, \& {Nanni}}]{Bressan2012}
{Bressan}, A., {Marigo}, P., {Girardi}, L., {et~al.} 2012, \mnras, 427, 127

\bibitem[{{Brogaard} {et~al.}(2011){Brogaard}, {Bruntt}, {Grundahl}, {Clausen},
  {Frandsen}, {Vandenberg}, \& {Bedin}}]{Brogaard2011}
{Brogaard}, K., {Bruntt}, H., {Grundahl}, F., {et~al.} 2011, \aap, 525, A2

\bibitem[{{Brogaard} {et~al.}(2018{\natexlab{a}}){Brogaard}, {Christiansen},
  {Grundahl}, {Miglio}, {Izzard}, {Tauris}, {Sand quist}, {VandenBerg},
  {Jessen-Hansen}, {Arentoft}, {Bruntt}, {Frandsen}, {Orosz}, {Feiden},
  {Mathieu}, {Geller}, {Shetrone}, {Ryde}, {Stello}, {Platais}, \&
  {Meibom}}]{Brogaard2018}
{Brogaard}, K., {Christiansen}, S.~M., {Grundahl}, F., {et~al.}
  2018{\natexlab{a}}, \mnras, 481, 5062

\bibitem[{{Brogaard} {et~al.}(2018{\natexlab{b}}){Brogaard}, {Hansen},
  {Miglio}, {Slumstrup}, {Frandsen}, {Jessen-Hansen}, {Lund}, {Bossini},
  {Thygesen}, {Davies}, {Chaplin}, {Arentoft}, {Bruntt}, {Grundahl}, \&
  {Handberg}}]{Brogaard2018A}
{Brogaard}, K., {Hansen}, C.~J., {Miglio}, A., {et~al.} 2018{\natexlab{b}},
  \mnras, 476, 3729

\bibitem[{{Brogaard} {et~al.}(2017){Brogaard}, {VandenBerg}, {Bedin}, {Milone},
  {Thygesen}, \& {Grundahl}}]{Brogaard2017}
{Brogaard}, K., {VandenBerg}, D.~A., {Bedin}, L.~R., {et~al.} 2017, \mnras,
  468, 645

\bibitem[{{Brogaard} {et~al.}(2012){Brogaard}, {VandenBerg}, {Bruntt},
  {Grundahl}, {Frandsen}, {Bedin}, {Milone}, {Dotter}, {Feiden}, {Stetson},
  {Sandquist}, {Miglio}, {Stello}, \& {Jessen-Hansen}}]{Brogaard2012}
{Brogaard}, K., {VandenBerg}, D.~A., {Bruntt}, H., {et~al.} 2012, \aap, 543,
  A106

\bibitem[{{Bruntt} {et~al.}(2010){Bruntt}, {Bedding}, {Quirion}, {Lo Curto},
  {Carrier}, {Smalley}, {Dall}, {Arentoft}, {Bazot}, \& {Butler}}]{Bruntt2010}
{Bruntt}, H., {Bedding}, T.~R., {Quirion}, P.~O., {et~al.} 2010, \mnras, 405,
  1907

\bibitem[{{Caffau} {et~al.}(2011){Caffau}, {Ludwig}, {Steffen}, {Freytag}, \&
  {Bonifacio}}]{Caffau2011}
{Caffau}, E., {Ludwig}, H.~G., {Steffen}, M., {Freytag}, B., \& {Bonifacio}, P.
  2011, \solphys, 268, 255

\bibitem[{{Casagrande} \& {VandenBerg}(2014)}]{Casagrande2014}
{Casagrande}, L. \& {VandenBerg}, D.~A. 2014, \mnras, 444, 392

\bibitem[{{Casagrande} \& {VandenBerg}(2018)}]{Casagrande2018}
{Casagrande}, L. \& {VandenBerg}, D.~A. 2018, \mnras, 479, L102

\bibitem[{{Chen} {et~al.}(2014){Chen}, {Girardi}, {Bressan}, {Marigo},
  {Barbieri}, \& {Kong}}]{Chen2014}
{Chen}, Y., {Girardi}, L., {Bressan}, A., {et~al.} 2014, \mnras, 444, 2525

\bibitem[{{Choi} {et~al.}(2016){Choi}, {Dotter}, {Conroy}, {Cantiello},
  {Paxton}, \& {Johnson}}]{Choi2016}
{Choi}, J., {Dotter}, A., {Conroy}, C., {et~al.} 2016, \apj, 823, 102

\bibitem[{{Claret}(2000)}]{Claret2000}
{Claret}, A. 2000, \aap, 363, 1081

\bibitem[{{Claret} \& {Bloemen}(2011)}]{Claret2011}
{Claret}, A. \& {Bloemen}, S. 2011, \aap, 529, A75

\bibitem[{{Coelho} {et~al.}(2005){Coelho}, {Barbuy}, {Mel{\'e}ndez},
  {Schiavon}, \& {Castilho}}]{Coelho2005}
{Coelho}, P., {Barbuy}, B., {Mel{\'e}ndez}, J., {Schiavon}, R.~P., \&
  {Castilho}, B.~V. 2005, \aap, 443, 735

\bibitem[{{Dotter}(2016)}]{Dotter2016}
{Dotter}, A. 2016, \apjs, 222, 8

\bibitem[{{Douglas} {et~al.}(2014){Douglas}, {Ag{\"u}eros}, {Covey}, {Bowsher},
  {Bochanski}, {Cargile}, {Kraus}, {Law}, {Lemonias}, {Arce}, {Fierroz}, \&
  {Kundert}}]{Douglas14}
{Douglas}, S.~T., {Ag{\"u}eros}, M.~A., {Covey}, K.~R., {et~al.} 2014, \apj,
  795, 161

\bibitem[{{Droege} {et~al.}(1997){Droege}, {Albertson}, {Gombert},
  {Gutzwiller}, {Molhant}, {Johnson}, {Skvarc}, {Wickersham}, {Richmond},
  {Rybski}, {Henden}, {Beser}, {Pittinger}, \& {Kluga}}]{TASS1997}
{Droege}, T.~F., {Albertson}, C., {Gombert}, G., {et~al.} 1997, in American
  Astronomical Society Meeting Abstracts, Vol. 190, American Astronomical
  Society Meeting Abstracts \#190, 30.10

\bibitem[{{El-Badry} {et~al.}(2018){El-Badry}, {Rix}, \&
  {Weisz}}]{El-Badry2018}
{El-Badry}, K., {Rix}, H.-W., \& {Weisz}, D.~R. 2018, \apjl, 860, L17

\bibitem[{{Etzel}(1981)}]{Etzel1981}
{Etzel}, P.~B. 1981, in Photometric and Spectroscopic Binary Systems, 111

\bibitem[{{Evans} {et~al.}(2018){Evans}, {Riello}, {De Angeli}, {Carrasco},
  {Montegriffo}, {Fabricius}, {Jordi}, {Palaversa}, {Diener}, {Busso},
  {Cacciari}, {van Leeuwen}, {Burgess}, {Davidson}, {Harrison}, {Hodgkin},
  {Pancino}, {Richards}, {Altavilla}, {Balaguer-N{\'u}{\~n}ez}, {Barstow},
  {Bellazzini}, {Brown}, {Castellani}, {Cocozza}, {De Luise}, {Delgado},
  {Ducourant}, {Galleti}, {Gilmore}, {Giuffrida}, {Holl}, {Kewley}, {Koposov},
  {Marinoni}, {Marrese}, {Osborne}, {Piersimoni}, {Portell}, {Pulone},
  {Ragaini}, {Sanna}, {Terrett}, {Walton}, {Wevers}, \&
  {Wyrzykowski}}]{Evans2018}
{Evans}, D.~W., {Riello}, M., {De Angeli}, F., {et~al.} 2018, \aap, 616, A4

\bibitem[{{Frandsen} {et~al.}(2018){Frandsen}, {Fredslund Andersen},
  {Brogaard}, {Jiang}, {Arentoft}, {Grundahl}, {Kjeldsen},
  {Christensen-Dalsgaard}, {Weiss}, {Pall{\'e}}, {Antoci}, {Kj{\ae}rgaard},
  {S{\o}rensen}, {Skottfelt}, \& {J{\o}rgensen}}]{Frandsen2018}
{Frandsen}, S., {Fredslund Andersen}, M., {Brogaard}, K., {et~al.} 2018, \aap,
  613, A53

\bibitem[{{Fredslund Andersen} {et~al.}(2019){Fredslund Andersen}, {Handberg},
  {Weiss}, {Frand sen}, {Sim{\'o}n-D{\'\i}az}, {Grundahl}, \&
  {Pall{\'e}}}]{Fredslund2019}
{Fredslund Andersen}, M., {Handberg}, R., {Weiss}, E., {et~al.} 2019, \pasp,
  131, 045003

\bibitem[{{Gaia Collaboration}(2018)}]{Gaia2018}
{Gaia Collaboration}. 2018, VizieR Online Data Catalog, I/345

\bibitem[{{Gaia Collaboration} {et~al.}(2017){Gaia Collaboration}, {van
  Leeuwen}, {Vallenari}, {Jordi}, {Lindegren}, {Bastian}, {Prusti}, {de
  Bruijne}, {Brown}, {Babusiaux}, \& et~al.}]{Gaia17}
{Gaia Collaboration}, {van Leeuwen}, F., {Vallenari}, A., {et~al.} 2017, \aap,
  601, A19

\bibitem[{{Gonz{\'a}lez} \& {Levato}(2006)}]{Gonzalez2006}
{Gonz{\'a}lez}, J.~F. \& {Levato}, H. 2006, \aap, 448, 283

\bibitem[{{Gossage} {et~al.}(2018){Gossage}, {Conroy}, {Dotter}, {Choi},
  {Rosenfield}, {Cargile}, \& {Dolphin}}]{Gossage18}
{Gossage}, S., {Conroy}, C., {Dotter}, A., {et~al.} 2018, \apj, 863, 67

\bibitem[{{Gray}(2009)}]{Gray2009}
{Gray}, D.~F. 2009, \apj, 697, 1032

\bibitem[{{Gray} \& {Martinez}(2019)}]{Gray2019}
{Gray}, D.~F. \& {Martinez}, A. 2019, \aj, 157, 92

\bibitem[{{Grevesse} \& {Noels}(1993)}]{Grevesse1993}
{Grevesse}, N. \& {Noels}, A. 1993, Physica Scripta Volume T, 47, 133

\bibitem[{{Griffin}(2012)}]{Griffin12}
{Griffin}, R.~F. 2012, Journal of Astrophysics and Astronomy, 33, 29

\bibitem[{{Griffin} {et~al.}(1988){Griffin}, {Gunn}, {Zimmerman}, \&
  {Griffin}}]{Griffin88}
{Griffin}, R.~F., {Gunn}, J.~E., {Zimmerman}, B.~A., \& {Griffin}, R.~E.~M.
  1988, \aj, 96, 172

\bibitem[{{Grundahl} {et~al.}(2017){Grundahl}, {Fredslund Andersen},
  {Christensen-Dalsgaard}, {Antoci}, {Kjeldsen}, {Hand berg}, {Houdek},
  {Bedding}, {Pall{\'e}}, {Jessen-Hansen}, {Silva Aguirre}, {White}, {Frand
  sen}, {Albrecht}, {Andersen}, {Arentoft}, {Brogaard}, {Chaplin},
  {Harps{\o}e}, {J{\o}rgensen}, {Karovicova}, {Karoff}, {Kj{\ae}rgaard
  Rasmussen}, {Lund}, {Sloth Lundkvist}, {Skottfelt}, {Norup S{\o}rensen},
  {Tronsgaard}, \& {Weiss}}]{Grundahl2017}
{Grundahl}, F., {Fredslund Andersen}, M., {Christensen-Dalsgaard}, J., {et~al.}
  2017, \apj, 836, 142

\bibitem[{{Handberg} {et~al.}(2017){Handberg}, {Brogaard}, {Miglio}, {Bossini},
  {Elsworth}, {Slumstrup}, {Davies}, \& {Chaplin}}]{Handberg2017}
{Handberg}, R., {Brogaard}, K., {Miglio}, A., {et~al.} 2017, \mnras, 472, 979

\bibitem[{{H{\o}g} {et~al.}(2000){H{\o}g}, {Fabricius}, {Makarov}, {Urban},
  {Corbin}, {Wycoff}, {Bastian}, {Schwekendiek}, \& {Wicenec}}]{Tycho-2-2000}
{H{\o}g}, E., {Fabricius}, C., {Makarov}, V.~V., {et~al.} 2000, \aap, 355, L27

\bibitem[{{Hroch}(2014)}]{Muniwin14}
{Hroch}, F. 2014, {Munipack: General astronomical image processing software},
  Astrophysics Source Code Library

\bibitem[{{Jurgenson} {et~al.}(2016){Jurgenson}, {Fischer}, {McCracken},
  {Sawyer}, {Giguere}, {Szymkowiak}, {Santoro}, \& {Muller}}]{Jurgenson2016}
{Jurgenson}, C., {Fischer}, D., {McCracken}, T., {et~al.} 2016, Journal of
  Astronomical Instrumentation, 5, 1650003

\bibitem[{{Jurgenson} {et~al.}(2014){Jurgenson}, {Fischer}, {McCracken},
  {Stoll}, {Szymkowiak}, {Giguere}, {Santoro}, \& {Muller}}]{Jurgenson2014}
{Jurgenson}, C.~A., {Fischer}, D.~A., {McCracken}, T.~M., {et~al.} 2014, in
  \procspie, Vol. 9147, Ground-based and Airborne Instrumentation for Astronomy
  V, 91477F

\bibitem[{{Khan} {et~al.}(2019){Khan}, {Miglio}, {Mosser}, {Arenou},
  {Belkacem}, {Brown}, {Katz}, {Casagrand e}, {Chaplin}, {Davies}, {Rendle},
  {Rodrigues}, {Bossini}, {Cantat-Gaudin}, {Elsworth}, {Girardi}, {North}, \&
  {Vallenari}}]{Khan2019}
{Khan}, S., {Miglio}, A., {Mosser}, B., {et~al.} 2019, \aap, 628, A35

\bibitem[{{Kharchenko}(2001)}]{Kharchenko2001}
{Kharchenko}, N.~V. 2001, Kinematika i Fizika Nebesnykh Tel, 17, 409

\bibitem[{{Lebreton} {et~al.}(2001){Lebreton}, {Fernandes}, \&
  {Lejeune}}]{Lebreton2001}
{Lebreton}, Y., {Fernandes}, J., \& {Lejeune}, T. 2001, \aap, 374, 540

\bibitem[{{Lodieu} {et~al.}(2019){Lodieu}, {Smart}, {P{\'e}rez-Garrido}, \&
  {Silvotti}}]{Lodieu2019}
{Lodieu}, N., {Smart}, R.~L., {P{\'e}rez-Garrido}, A., \& {Silvotti}, R. 2019,
  \aap, 623, A35

\bibitem[{{Mamajek} {et~al.}(2002){Mamajek}, {Meyer}, \&
  {Liebert}}]{Mamajek2002}
{Mamajek}, E.~E., {Meyer}, M.~R., \& {Liebert}, J. 2002, \aj, 124, 1670

\bibitem[{{Mamajek} {et~al.}(2006){Mamajek}, {Meyer}, \&
  {Liebert}}]{Mamajek2006}
{Mamajek}, E.~E., {Meyer}, M.~R., \& {Liebert}, J. 2006, \aj, 131, 2360

\bibitem[{{Mart{\'\i}n} {et~al.}(2018){Mart{\'\i}n}, {Lodieu}, {Pavlenko}, \&
  {B{\'e}jar}}]{Martin18}
{Mart{\'\i}n}, E.~L., {Lodieu}, N., {Pavlenko}, Y., \& {B{\'e}jar}, V. J.~S.
  2018, \apj, 856, 40

\bibitem[{{McClure}(1982)}]{McClure82}
{McClure}, R.~D. 1982, \apj, 254, 606

\bibitem[{{Mellon} {et~al.}(2019){Mellon}, {Mamajek}, {Stuik}, {Zwintz},
  {Kenworthy}, {Talens}, {Burggraaff}, {Bailey}, {Dorval}, {Lomberg}, {Kuhn},
  \& {Ireland}}]{Mellon2019}
{Mellon}, S.~N., {Mamajek}, E.~E., {Stuik}, R., {et~al.} 2019, \apjs, 244, 15

\bibitem[{{Miglio} {et~al.}(2012){Miglio}, {Brogaard}, {Stello}, {Chaplin},
  {D'Antona}, {Montalb{\'a}n}, {Basu}, {Bressan}, {Grundahl}, {Pinsonneault},
  {Serenelli}, {Elsworth}, {Hekker}, {Kallinger}, {Mosser}, {Ventura},
  {Bonanno}, {Noels}, {Silva Aguirre}, {Szabo}, {Li}, {McCauliff}, {Middour},
  \& {Kjeldsen}}]{Miglio2012}
{Miglio}, A., {Brogaard}, K., {Stello}, D., {et~al.} 2012, \mnras, 419, 2077

\bibitem[{{Miglio} {et~al.}(2016){Miglio}, {Chaplin}, {Brogaard}, {Lund},
  {Mosser}, {Davies}, {Handberg}, {Milone}, {Marino}, {Bossini}, {Elsworth},
  {Grundahl}, {Arentoft}, {Bedin}, {Campante}, {Jessen-Hansen}, {Jones},
  {Kuszlewicz}, {Malavolta}, {Nascimbeni}, \& {Sandquist}}]{Miglio2016}
{Miglio}, A., {Chaplin}, W.~J., {Brogaard}, K., {et~al.} 2016, \mnras, 461, 760

\bibitem[{{Miglio} {et~al.}(2020){Miglio}, {Chiappini}, {Mackereth}, {Davies},
  {Brogaard}, {Casagrande}, {Chaplin}, {Girardi}, {Kawata}, {Khan}, {Izzard},
  {Montalban}, {Mosser}, {Vincenzo}, {Bossini}, {Noels}, {Rodrigues},
  {Valentini}, \& {Mand el}}]{Miglio2020}
{Miglio}, A., {Chiappini}, C., {Mackereth}, T., {et~al.} 2020, arXiv e-prints,
  arXiv:2004.14806

\bibitem[{{Paxton} {et~al.}(2011){Paxton}, {Bildsten}, {Dotter}, {Herwig},
  {Lesaffre}, \& {Timmes}}]{Paxton2011}
{Paxton}, B., {Bildsten}, L., {Dotter}, A., {et~al.} 2011, \apjs, 192, 3

\bibitem[{{Paxton} {et~al.}(2018){Paxton}, {Schwab}, {Bauer}, {Bildsten},
  {Blinnikov}, {Duffell}, {Farmer}, {Goldberg}, {Marchant}, {Sorokina},
  {Thoul}, {Townsend}, \& {Timmes}}]{Paxton2018}
{Paxton}, B., {Schwab}, J., {Bauer}, E.~B., {et~al.} 2018, \apjs, 234, 34

\bibitem[{{Perryman} {et~al.}(1998){Perryman}, {Brown}, {Lebreton}, {Gomez},
  {Turon}, {Cayrel de Strobel}, {Mermilliod}, {Robichon}, {Kovalevsky}, \&
  {Crifo}}]{Perryman98}
{Perryman}, M.~A.~C., {Brown}, A.~G.~A., {Lebreton}, Y., {et~al.} 1998, \aap,
  331, 81

\bibitem[{{Peterson} \& {Solensky}(1987)}]{Peterson87}
{Peterson}, D.~M. \& {Solensky}, R. 1987, \apj, 315, 286

\bibitem[{{Peterson} \& {Solensky}(1988)}]{Peterson88}
{Peterson}, D.~M. \& {Solensky}, R. 1988, \apj, 333, 256

\bibitem[{{Popper} \& {Etzel}(1981)}]{Popper1981}
{Popper}, D.~M. \& {Etzel}, P.~B. 1981, \aj, 86, 102

\bibitem[{{Rodrigues} {et~al.}(2017){Rodrigues}, {Bossini}, {Miglio},
  {Girardi}, {Montalb{\'a}n}, {Noels}, {Trabucchi}, {Coelho}, \&
  {Marigo}}]{Rodrigues2017}
{Rodrigues}, T.~S., {Bossini}, D., {Miglio}, A., {et~al.} 2017, \mnras, 467,
  1433

\bibitem[{{Rucinski}(1999)}]{Rucinski1999}
{Rucinski}, S. 1999, Turkish Journal of Physics, 23, 271

\bibitem[{{Rucinski}(2002)}]{Rucinski2002}
{Rucinski}, S.~M. 2002, \aj, 124, 1746

\bibitem[{{Salaris} \& {Bedin}(2018)}]{Salaris2018}
{Salaris}, M. \& {Bedin}, L.~R. 2018, \mnras, 480, 3170

\bibitem[{{Sandquist} {et~al.}(2016){Sandquist}, {Jessen-Hansen}, {Shetrone},
  {Brogaard}, {Meibom}, {Leitner}, {Stello}, {Bruntt}, {Antoci}, {Orosz},
  {Grundahl}, \& {Frandsen}}]{Sandquist2016}
{Sandquist}, E.~L., {Jessen-Hansen}, J., {Shetrone}, M.~D., {et~al.} 2016,
  \apj, 831, 11

\bibitem[{{Sandquist} {et~al.}(2020){Sandquist}, {Stello}, {Arentoft},
  {Brogaard}, {Grundahl}, {Vanderburg}, {Hedlund}, {DeWitt}, {Ackerman},
  {Aguilar}, {Buckner}, {Juarez}, {Ortiz}, {Richarte}, {Rivera}, \&
  {Schlapfer}}]{Sandquist2020}
{Sandquist}, E.~L., {Stello}, D., {Arentoft}, T., {et~al.} 2020, \aj, 159, 96

\bibitem[{{Schiller} \& {Milone}(1987)}]{Schiller87}
{Schiller}, S.~J. \& {Milone}, E.~F. 1987, \aj, 93, 1471

\bibitem[{{Schr{\"o}der} {et~al.}(2020){Schr{\"o}der}, {Mittag}, {Jack},
  {Rodr{\'\i}guez Jim{\'e}nez}, \& {Schmitt}}]{Schroder2020}
{Schr{\"o}der}, K.~P., {Mittag}, M., {Jack}, D., {Rodr{\'\i}guez Jim{\'e}nez},
  A., \& {Schmitt}, J.~H.~M.~M. 2020, \mnras, 492, 1110

\bibitem[{{Schwan}(1991)}]{Schwan91}
{Schwan}, H. 1991, \aap, 243, 386

\bibitem[{{Slumstrup} {et~al.}(2019){Slumstrup}, {Grundahl}, {Silva Aguirre},
  \& {Brogaard}}]{Slumstrup2019}
{Slumstrup}, D., {Grundahl}, F., {Silva Aguirre}, V., \& {Brogaard}, K. 2019,
  \aap, 622, A111

\bibitem[{{Southworth}(2013)}]{Southworth2013}
{Southworth}, J. 2013, \aap, 557, A119

\bibitem[{{Southworth}(2015)}]{Southworth2015}
{Southworth}, J. 2015, {JKTLD: Limb darkening coefficients}

\bibitem[{{Southworth} {et~al.}(2007){Southworth}, {Bruntt}, \&
  {Buzasi}}]{Southworth2007}
{Southworth}, J., {Bruntt}, H., \& {Buzasi}, D.~L. 2007, \aap, 467, 1215

\bibitem[{{Southworth} {et~al.}(2004){Southworth}, {Maxted}, \&
  {Smalley}}]{Southworth2004}
{Southworth}, J., {Maxted}, P.~F.~L., \& {Smalley}, B. 2004, \mnras, 351, 1277

\bibitem[{{Stetson}(1987)}]{Daophot87}
{Stetson}, P.~B. 1987, \pasp, 99, 191

\bibitem[{{Svechnikov} \& {Perevozkina}(2004)}]{Svechnikov04}
{Svechnikov}, M.~A. \& {Perevozkina}, E.~L. 2004, VizieR Online Data Catalog,
  5121, 5

\bibitem[{{Takeda} \& {Honda}(2020)}]{Takeda2020}
{Takeda}, Y. \& {Honda}, S. 2020, \aj, 159, 174

\bibitem[{{Talens} {et~al.}(2018){Talens}, {Deul}, {Stuik}, {Burggraaff},
  {Lesage}, {Spronck}, {Mellon}, {Bailey}, {Mamajek}, {Kenworthy}, \&
  {Snellen}}]{Talens2018}
{Talens}, G.~J.~J., {Deul}, E.~R., {Stuik}, R., {et~al.} 2018, \aap, 619, A154

\bibitem[{{Talens} {et~al.}(2017){Talens}, {Spronck}, {Lesage}, {Otten},
  {Stuik}, {Pollacco}, \& {Snellen}}]{Talens2017}
{Talens}, G.~J.~J., {Spronck}, J.~F.~P., {Lesage}, A.~L., {et~al.} 2017, \aap,
  601, A11

\bibitem[{{Torres}(2010)}]{Torres2010}
{Torres}, G. 2010, \aj, 140, 1158

\bibitem[{{Torres}(2019)}]{Torres2019B}
{Torres}, G. 2019, \apj, 883, 105

\bibitem[{{Torres} \& {Ribas}(2002)}]{Torres02}
{Torres}, G. \& {Ribas}, I. 2002, \apj, 567, 1140

\bibitem[{{Torres} {et~al.}(2019){Torres}, {Stefanik}, \&
  {Latham}}]{Torres2019A}
{Torres}, G., {Stefanik}, R.~P., \& {Latham}, D.~W. 2019, \apj, 885, 9

\bibitem[{{van Leeuwen}(2007)}]{vanLeeuwen2007}
{van Leeuwen}, F. 2007, \aap, 474, 653

\bibitem[{{VandenBerg} {et~al.}(2014){VandenBerg}, {Bergbusch}, {Ferguson}, \&
  {Edvardsson}}]{VandenBerg2014}
{VandenBerg}, D.~A., {Bergbusch}, P.~A., {Ferguson}, J.~W., \& {Edvardsson}, B.
  2014, \apj, 794, 72

\bibitem[{{Watson} {et~al.}(2006){Watson}, {Henden}, \& {Price}}]{Watson06}
{Watson}, C.~L., {Henden}, A.~A., \& {Price}, A. 2006, Society for Astronomical
  Sciences Annual Symposium, 25, 47

\end{thebibliography}

\appendix
\section{Theoretical corrections to the astroseismic scaling relations.}
For reference, we add here the theoretical corrections to $\Delta \nu$ in the asteroseismic scaling relations that we used. These originate from the work of \citet{Rodrigues2017}, which however only shows the correction as a function of $T_{\rm eff}$. Here we include the corresponding diagrams for the $\Delta \nu$ correction as a function of $\nu_{\rm max}$ (A. Miglio, priv. comm.).

\begin{figure}
   \centering
    \includegraphics[width=8cm]{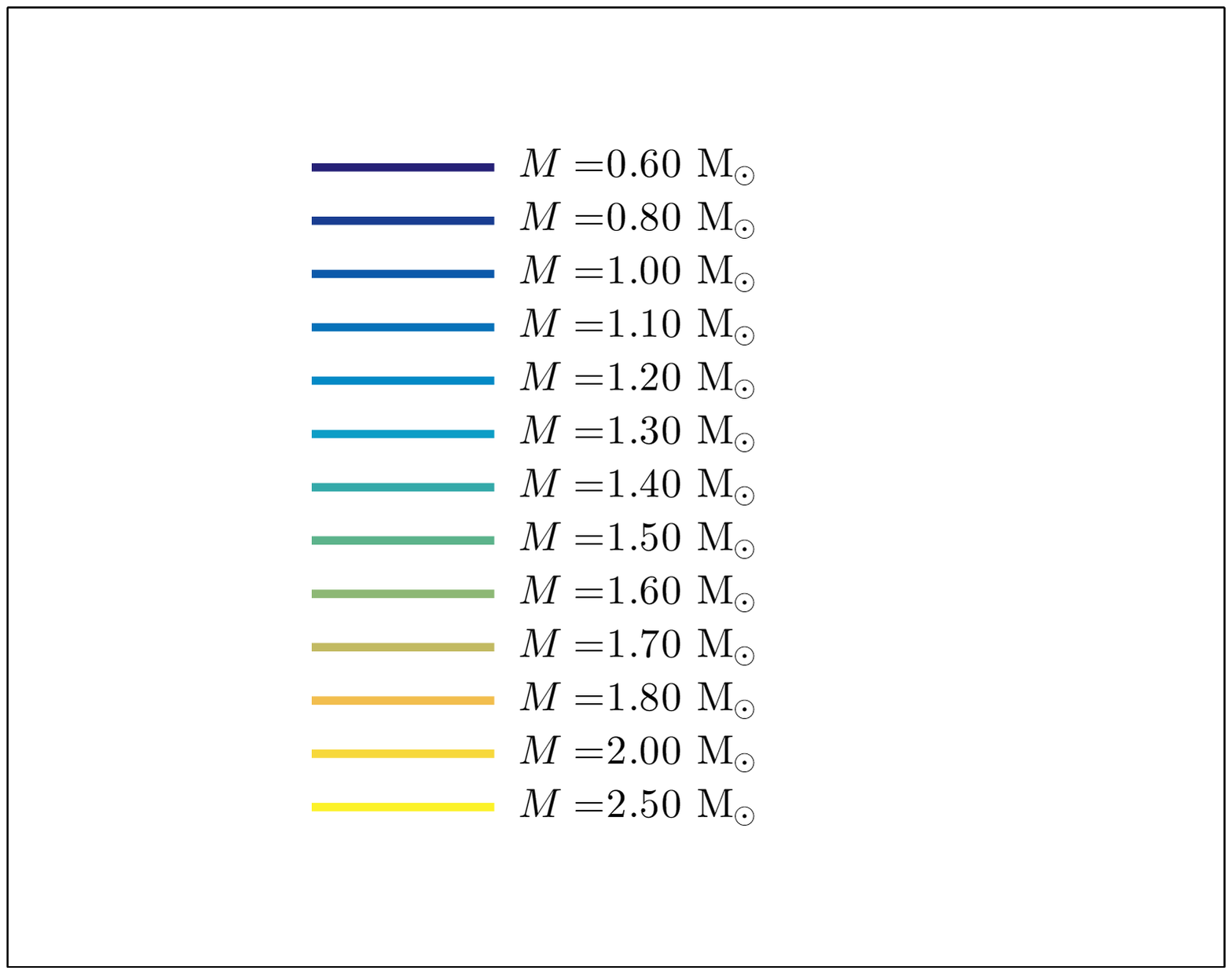}
    \includegraphics[width=\hsize]{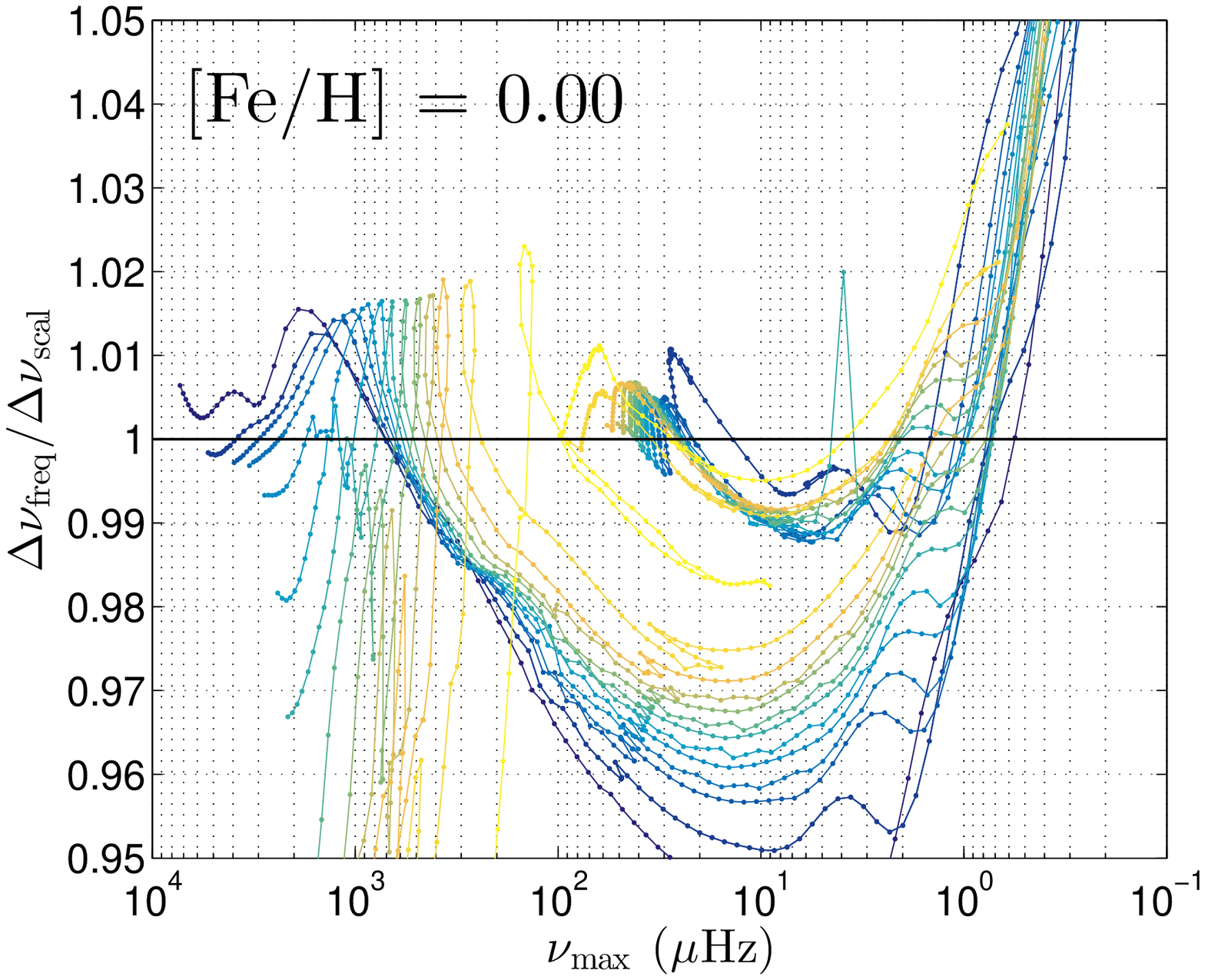}
    \includegraphics[width=\hsize]{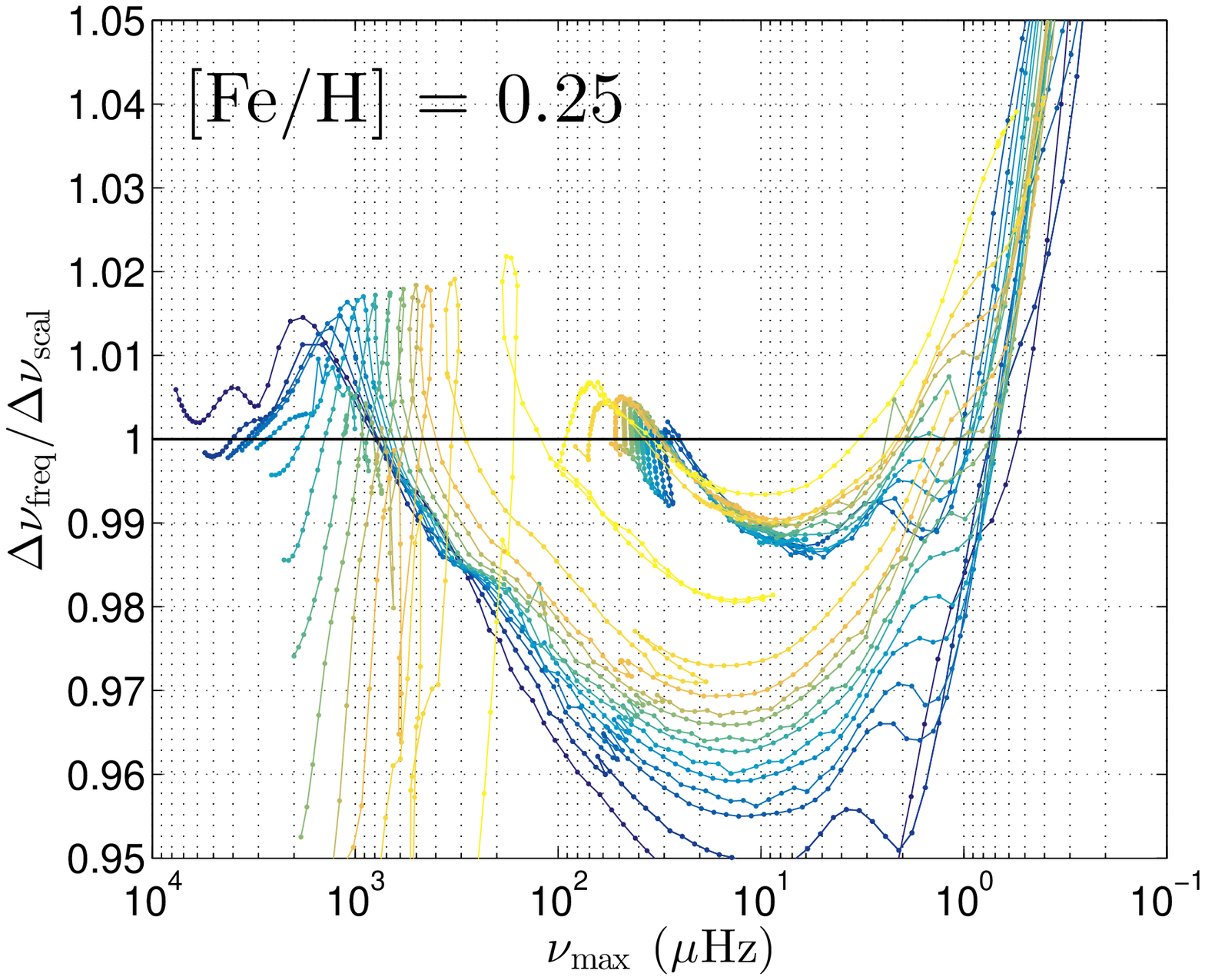}
      \caption{Theoretical corrections to $\Delta \nu$ in the asteroseismic scaling relations. See \citet{Rodrigues2017} for details.}
         \label{fig:scaling}
   \end{figure}

\end{document}